\documentclass[oneside, fleq , 11pt]{article}

\usepackage[dvips]{graphicx}
\usepackage{amsmath, comment}
\usepackage{amssymb}
\usepackage{sectsty}
\usepackage{titlesec}
\usepackage[margin=1in]{geometry}
\usepackage{fancyhdr}
\usepackage{color, colortbl}
\usepackage{amsthm}
\usepackage{xcolor,colortbl}
\usepackage{enumitem}
\usepackage{bm}
\usepackage{ulem}
\usepackage{graphicx}
\usepackage{subfig}

\definecolor{Gray}{gray}{0.85}
\definecolor{LightCyan}{rgb}{0.88,1,1}
\newcolumntype{a}{>{\columncolor{Gray}}c}

\theoremstyle{plain}

\newtheorem*{thm*}{Theorem}

\allsectionsfont{\bf \large }

\titlespacing*{\section}{0pt}{12pt}{3pt}{}
\titlespacing*{\subsection}{0pt}{12pt}{3pt}{}

\newcommand{\ben}{\begin{enumerate}}
\newcommand{\een}{\end{enumerate}}
\newcommand{\bit}{\begin{itemize}}
\newcommand{\eit}{\end{itemize}}

\definecolor{Gray}{gray}{0.9}
\newcolumntype{g}{>{\columncolor{Gray}}c}
\allowdisplaybreaks

\usepackage[square,numbers]{natbib}
\def \ts {{\,}}

\def \hbar {{\bar h}}

\newcommand*{\red}{\textcolor{red}}

\titleformat{\section}{\normalfont\normalsize \bfseries }{\thesection.}{0.5em}{}
\titlespacing*{\section}{0em}{1.0em}{-0.75em}

\begin{document}

\begin{center}{\Large ~Can Testing Ease Social Distancing Measures?
\break
Future Evolution of COVID-19 in NYC}
\\
\vspace{4mm}
\normalsize
Omar El Housni$^1$\footnote{Email addresses of the authors are 
oe2148@columbia.edu, 
ms3268@cornell.edu, 
rusmevic@marshall.usc.edu, 
topaloglu@orie.cornell.edu, and  
ziya@email.unc.edu.}\!\!, Mika Sumida$^2$, Paat Rusmevichientong$^3$, Huseyin Topaloglu$^2$, Serhan Ziya$^4$
\\
\vspace{2mm}
\footnotesize
$^1$Department of Industrial Engineering and Operations Research, Columbia University, New York, NY 10027
\\
$^2$School of Operations Research and Information Engineering, Cornell Tech, New York, NY 10044
\\
$^3$Marshall School of Business, University of Southern California, Los Angeles, CA 90089
\\
$^4$Department of Statistics and Operations Research, University of North Carolina, Chapel Hill, NC 27599
\\
\normalsize
\vspace{1mm}
\today
\end{center}

\vspace{-7mm}

\begin{abstract}
\noindent The \mbox{New York State on Pause} executive order came into effect on March 22 with the goal of ensuring adequate social distancing to alleviate the spread of COVID-19. Pause will remain effective in New York City in some form until early June. We use a compartmentalized model to study the effects of testing capacity and social distancing measures on the evolution of the pandemic in the \mbox{post-Pause} period in the City. We find that testing capacity must increase dramatically if it is to counterbalance even relatively small relaxations in social distancing measures in the immediate post-Pause period. In particular, if the City performs 20,000 tests per day and relaxes the social distancing measures to the pre-Pause norms, then the total number of deaths by the end of September can reach 250,000.~By keeping the social distancing measures to somewhere halfway between the pre- and in-Pause norms and performing 100,000 tests per day, the total number of deaths by the end of September can be kept at around 27,000. Going back to the pre-Pause social distancing norms quickly must be accompanied by an exorbitant testing capacity, if one is to suppress excessive deaths. If the City is to go back to the \mbox{pre-Pause} social distancing norms in the immediate post-Pause period and keep the total number of deaths by the end of September at around 35,000, then it should be performing 500,000 tests per day. Our findings have important implications on the magnitude of the testing capacity the City needs as it relaxes the social distancing measures to reopen its economy.

\end{abstract}

\section{Background and Overview}
\label{sec:mainbody}
In this study, we analyze the impact of social distancing measures and testing capacity on the evolution of COVID-19 in New York City. On March 22, the New York State on Pause executive order came into effect to ensure adequate social distancing and it has been helpful in controlling the spread of the disease and preventing healthcare resources from stretching too thin. Pause will remain in effect in the City in some form until early June. 
Social distancing measures, to the degree they are practiced today, are not sustainable. Testing, particularly together with contact tracing, could help relax social distancing measures, facilitating the partial reopening of the economy while keeping the spread of the disease under control. However, the level of testing capacity necessary to allow noticeable relaxation in the social distancing norms and to help reopen the economy is not yet clear. We consider various possibilities regarding the social distancing measures and testing capacity availability in the immediate post-Pause period. We project the evolution of the pandemic in the City until the end of September. To make our projections, we use a compartmentalized model that captures the limited availability of testing capacity, while allowing some contact tracing to identify, test, and isolate individuals at risk.

We find that testing capacity must increase dramatically if it is to counterbalance somewhat small levels of relaxation in the social distancing measures in the post-Pause period.~If the social distancing measures in the post-Pause period are relaxed to pre-Pause norms and the City continues to perform 20,000 tests per day, then the number of deaths by the end of September can reach 250,000. If the social distancing norms are fully  relaxed, then even with a testing capacity of 100,000 per day, the City still faces a second wave of the pandemic that would likely supersede the one it experienced in April. By keeping the social distancing measures halfway between the pre- and in-Pause levels and performing 100,000 tests per day in the post-Pause period, the number of deaths by the end of September can be kept at around 27,000. If the City is to go back to pre-Pause social distancing norms and keep the number of deaths by the end of September under 35,000, then it should be performing 500,000 tests per day. The message is clear.  If one considers for a moment going back to the \mbox{pre-Pause} life quickly in the post-Pause period, then the City should be ramping up its testing capacity to 500,000, but even then, there will be a price to pay.

\section{Compartmentalized Model}
Our model follows the standard susceptible-infected-recovered paradigm with compartments capturing individuals classified along the dimensions of infected, noninfected, and recovered, as well as symptomatic-isolated, asymptomatic-isolated, and asymptomatic-nonisolated. Once an individual is tested positive, a certain number of individuals who are expected to have been in close contact with the positive individual are transferred to a certain isolated compartment. In this way, we build a contact tracing mechanism to ensure that those who have been in contact with a positive individual reduce their contact with the rest of the population.  Running our model with a starting date of March 2, we get a trajectory of the pandemic that is in reasonably close agreement with the actual trajectory in the City so far. In the appendix, we give a discussion of our model and provide comparisons of its output with the actual trajectory of the pandemic.

\section{A Close Look at Two Testing Scenarios}

\begin{figure}[t]
\begin{center}
20,000 tests per day
\\
\includegraphics[scale=0.57]{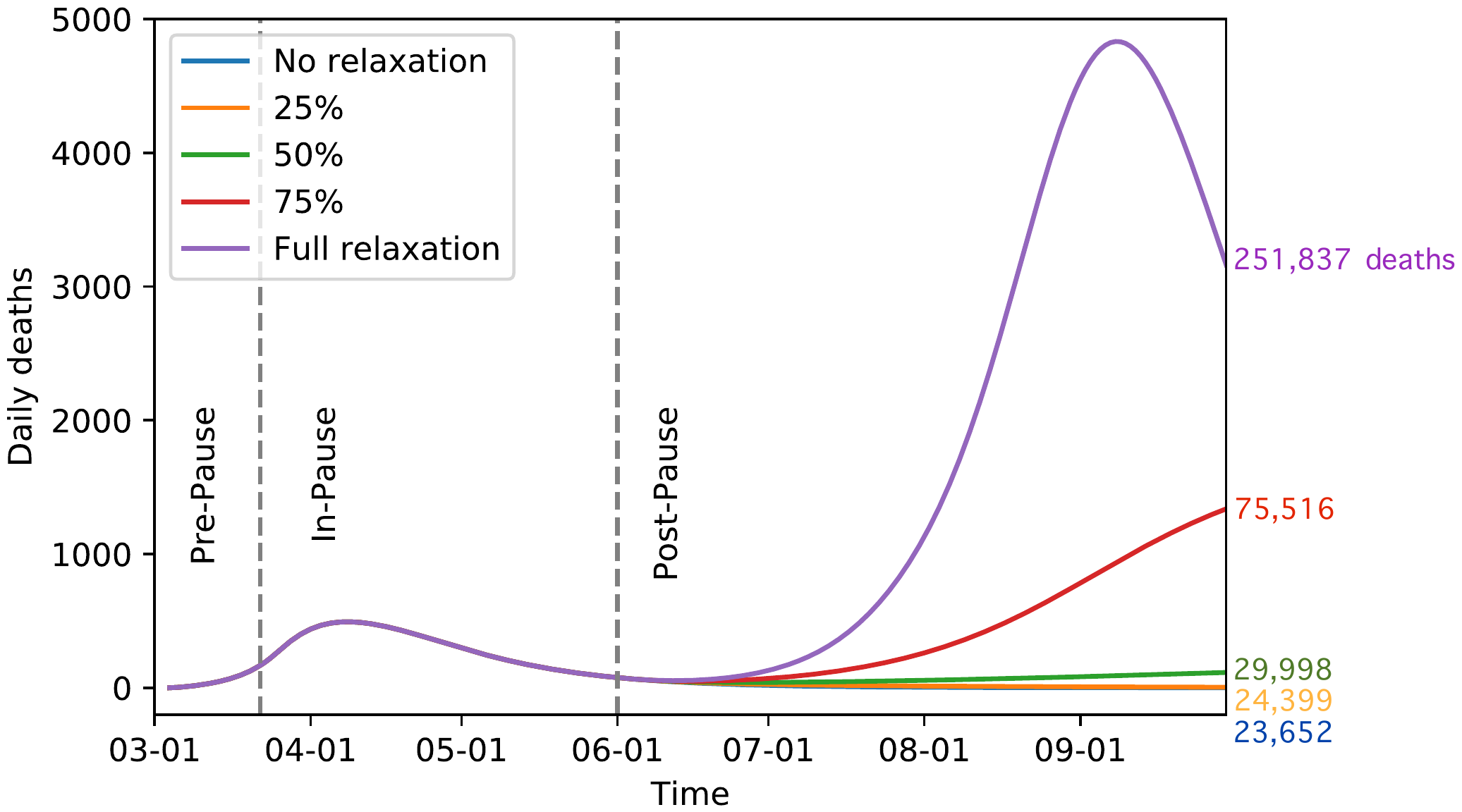}
\vspace{2mm}
\\
100,000 tests per day
\\
\includegraphics[scale=0.57]{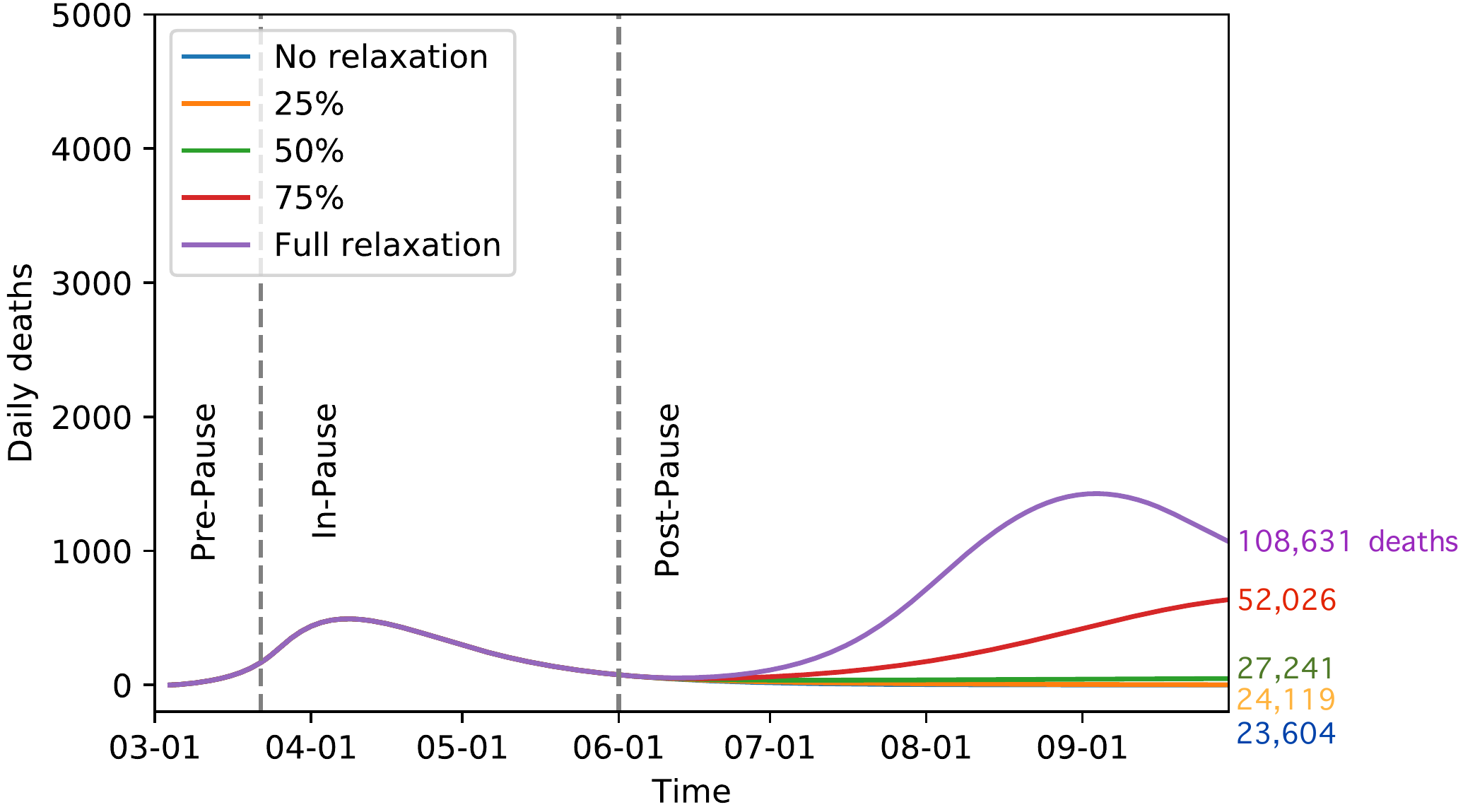}

\caption{Trajectory of the daily deaths under testing capacities of 20,000 and 100,000  per day.}
\vspace{-8mm}
\label{fig:traj}
\end{center}
\end{figure}

In the top and bottom panels of Figure \ref{fig:traj}, we give the trajectory of projected daily deaths under two testing capacity scenarios of 20,000 and 100,000 tests per day. Each data series in the two panels corresponds to a different level of social distancing practiced in the post-Pause period, expressed as a percentage relaxation in the \mbox{in-Pause} social distancing norms. Specifically, 0\% is no relaxation, corresponding to the strictest level of social distancing that was implemented in the in-Pause period, whereas 100\% is full relaxation, corresponding to no social distancing, as in life in the pre-Pause period.  Other percentages reflect different levels of relaxation the City may consider, each with a certain R-naught value between the two extremes. All \mbox{newly-imposed} social distancing measures come into effect on June 1, the beginning of the post-Pause period. The testing capacities of 20,000 and 100,000 tests per day represent the current situation and an order of magnitude bump.

Noting the purple data series in the figure, fully relaxing the social distancing measures immediately in the post-Pause period and going back to pre-Pause practices will lead to another wave peaking in late August, regardless of whether the City has 20,000 or 100,000 per day testing capacity. Testing capacity will surely have an impact on the size of the next wave. Under 20,000 per day testing capacity, full relaxation of social distancing measures will result in 252,000 deaths by the end of September. Note that the labels on the right end of each panel give the total number of deaths by the end of September. Even with 100,000 per day testing capacity, if the social distancing norms are fully relaxed in the post-Pause period, then the next wave in August will be significantly larger than what the City already experienced in April, ultimately resulting in 109,000 deaths by the end of September. %

Thus, quickly going back to the pre-Pause practices in the post-Pause period will likely have dramatic consequences even if the City can perform 100,000 tests per day. Consider following social distancing measures somewhere halfway between the pre- and in-Pause levels, corresponding to the green data series in the figure. In this case, if the City keeps on performing 20,000 tests per day in the \mbox{post-Pause} period, then the number of deaths by the end of September adds up to 30,000.~Increasing the testing capacity to 100,000 per day reduces the number of deaths by the end of September to 27,000, but even in this case, the daily deaths demonstrate a slightly increasing trend at the end of September.

Being more liberal with the social distancing measures in the post-Pause period and relaxing them even more than  halfway between the pre- and in-Pause levels will be problematic. Consider relaxing the  social distancing measures by 75\%, corresponding to the red data series in the figure. Under both testing capacities of 20,000 and 100,000 per day, a second wave picks up at the end of September. The testing capacity will affect the magnitude of the wave, but the number of deaths by the end of August reaches 52,000 even with a testing capacity of 100,000 per day.

In Figure \ref{fig:traj}, full (100\%) and no (0\%) relaxation in social distancing practices correspond to life before and after March 22. We emphasize that the interpretation of other levels of social distancing measures requires more care. In particular, the other levels of social distancing measures correspond to different R-naught values between the two extremes and they should \underline{\it not} be interpreted as a suggested level of occupancy in business establishments. There is work to be done to convert these social distancing levels into public health policy.

\section{What Level of Testing Capacity Allows for Reopening}

The discussion in the previous section indicates there are risks if the City quickly relaxes the social distancing norms in the post-Pause period. A natural question is what testing capacity would likely be sufficient to keep the risk of a second wave at bay while relaxing the social distancing measures to a degree that the economy can start functioning.

\begin{figure}[t]

\begin{center}
\includegraphics[scale=0.55]{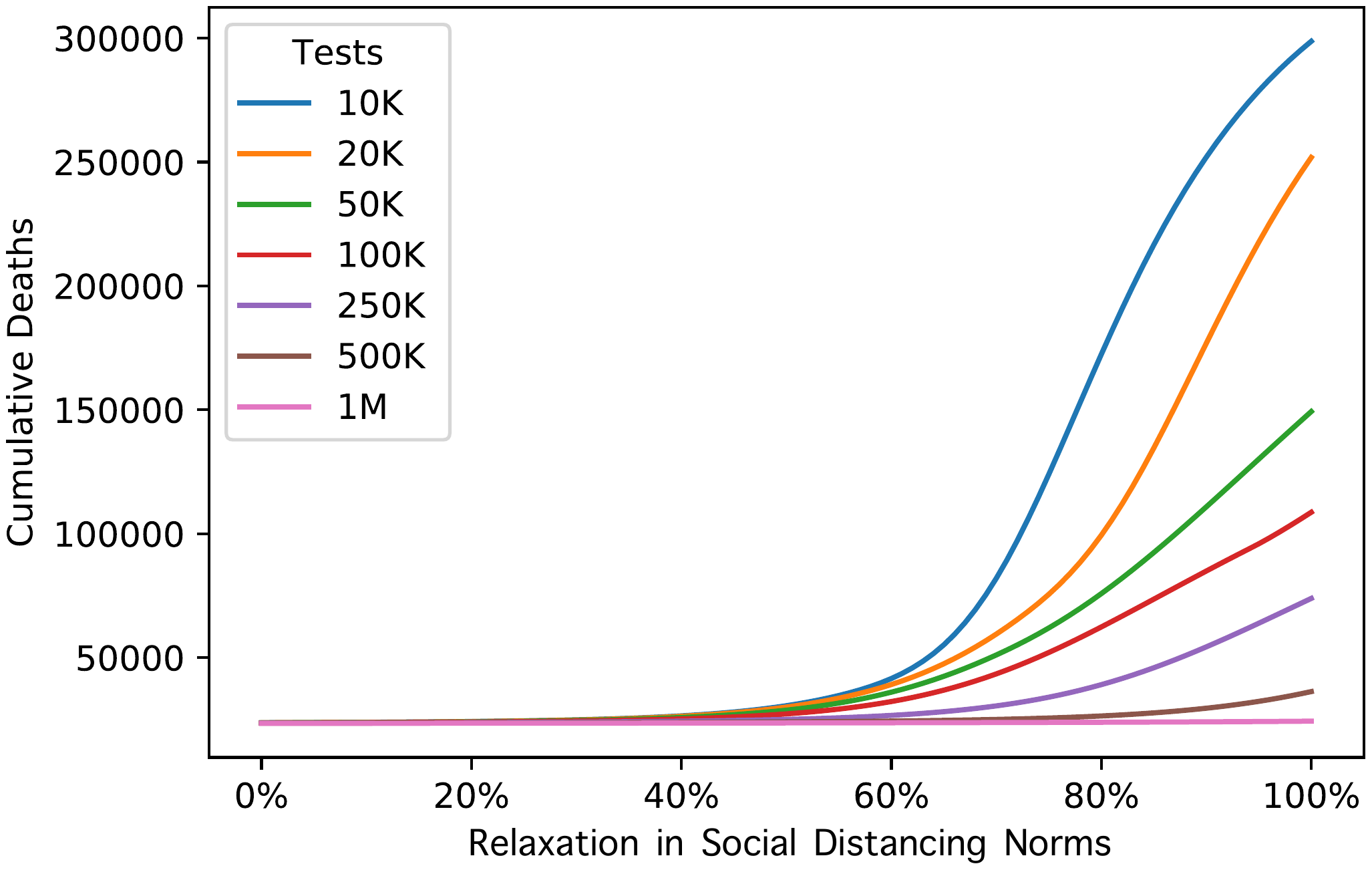}
\vspace{2mm}
\\
~\hspace{-2.5mm}
\includegraphics[scale=0.54]{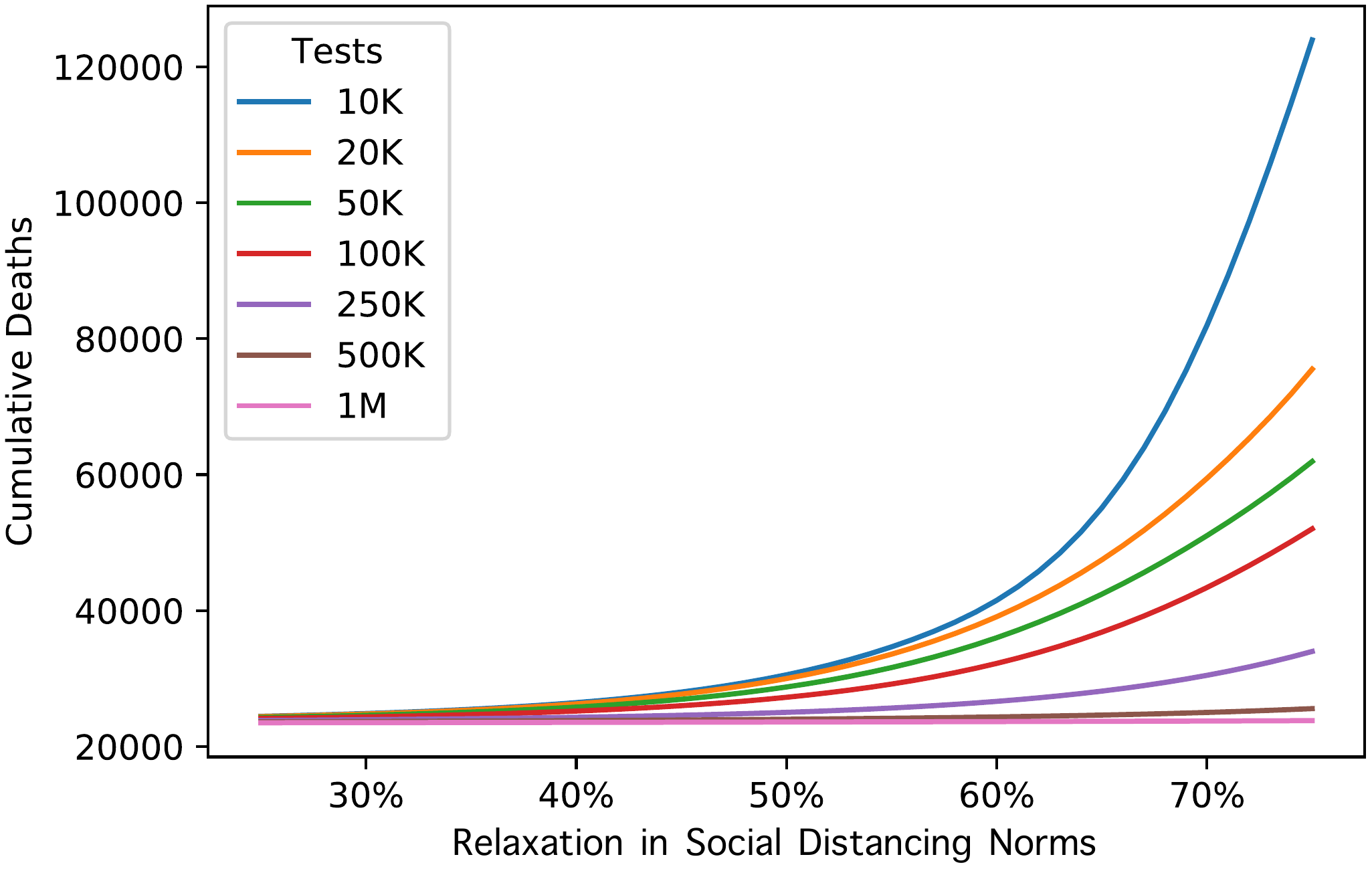}

\caption{Tradeoff between the deaths by the end of September and the social distancing norms.}
\vspace{-4mm}
\label{fig:front}
\end{center}
\end{figure}

In the top panel of Figure \ref{fig:front}, we show the tradeoff between the total number of deaths by the end of September and the level of social distancing practiced, under different testing capacities. The horizontal axis shows the level of social distancing, expressed, as before, as a percentage relaxation in the \mbox{in-Pause} social distancing norms. Thus, 0\% is no relaxation, corresponding to the strictest level of social distancing observed during Pause, whereas 100\% is full relaxation, corresponding to the loosest level of social distancing that was in the pre-Pause period. The vertical axis shows the total number of projected deaths by the end of September. Each data series corresponds to a different level of testing capacity. In the bottom panel of Figure \ref{fig:front}, we give the same information but zoom in on the more relevant region of social distancing measures. In our model, all new testing capacity and social distancing norms become effective on June 1. Before May 1, we use the actual number of tests performed by the City and extrapolate between early-May and June 1.

A good starting point is to consider the case where the City relaxes the social distancing measures halfway between the pre- and in-Pause norms, corresponding to 50\% on the horizontal axis in the figure. Keeping the social distancing measures any stricter will likely not be useful for reopening the economy. We observe that keeping a testing capacity of 20,000 per day results in a total of 30,000 deaths by the end of September. Raising the testing capacity to 100,000 per day reduces the number of deaths by the end of September to 27,000. Consider relaxing the in-Pause social distancing measures by 75\% during the post-Pause period.  Even by performing~250,000 tests per day, the total number of deaths by the end of September can only be kept at 34,000. Thus, if one  considers for a moment quickly going back to the pre-Pause life, then the City should be ramping up its testing capacity dramatically, up to 500,000 per day. Even then, there will be a price to pay and testing is far from being a perfect substitute for social distancing.

\section{Discussion and Next Steps}

The analysis in this study puts all new social distancing measures and testing capacities into effect immediately on June 1. In practice, the change will be gradual.  Our model provides a tool to investigate how to relax social distancing measures gradually as the City ramps up its testing capacity over time. An analysis along these lines will be our next line of attack. Moreover, as discussed in the appendix, we have a contact tracing mechanism in our model, where close contacts of every positive case are put into isolation. In other words, our model implicitly assumes that the bottleneck is the testing capacity, not the contract tracing capacity. We plan to use our model to estimate the number of contact tracers needed to keep up with the increasing testing capacity and new cases. The interpretation of the pre- and \mbox{in-Pause} social distancing measures are somewhat clear, each corresponding to life before and after March 22. Relaxing social distancing measures by any percentage corresponds to adjusting R-naught to lie appropriately between the R-naught values of these two extremes. It should be noted that it will require work to understand which public health policy would achieve what level of R-naught value. Lastly, we believe that the death projections from our model are on the optimistic side, because our model assumes the presence of a streamlined tracing mechanism and individuals in certain compartments are willing to isolate themselves, infecting others minimally. Such  optimistic behavior of our model should be taken into consideration when interpreting our results.

\bibliographystyle{abbrvnat}
\bibliography{references}

\newpage
\appendix

\appendix
\noindent {\bf \Large Appendix}
\\
The appendix is organized as follows. In Appendix \ref{section:model}, we discuss our model at a high level.~This section avoids any technical details but it is useful for seeing how our model works, what compartments it uses, and how the different compartments interact with each other. In Appendix~\ref{section:system-dynamics}, we provide the full mathematical details of our model. In Appendix \ref{section:model-parameters}, we elaborate on the parameters of our model and how they are estimated. We borrow some of the parameters from the literature and derive others based off of problem primitives. We explain how we arrive at the derived model parameters. In Appendix \ref{section:model-validation}, we compare the output of our model with the trajectory of the pandemic in the City until early May and demonstrate that our model does a reasonably good job of predicting the trajectory that has been observed so far.

\section{Model at a High Level}
\label{section:model}

\vspace{10pt}

In our model, we capture the two benefits of testing, which are isolating the specific individual tested positive and tracing the contacts of the positive individual. In particular, our model distinguishes the infected individuals who know that they are infected from the infected individuals who do not know whether they are infected. Once a person is tested positive, this person is an infected individual who knows that she is infected. Such a person minimally infects others. Moreover, considering the individuals who do not know whether they are infected, we classify them as symptomatic-isolated, \mbox{asymptomatic-isolated}, or \mbox{asymptomatic-nonisolated}. Once an individual is tested positive, a certain number of people around this person are classified as~\mbox{asymptomatic-isolated} people, so these individuals  reduce their contact with the general population, preventing them from infecting others to some extent.

\begin{figure}[t]
\begin{center}
~
\includegraphics[scale=0.46,trim=2cm 5cm 0cm 1cm]{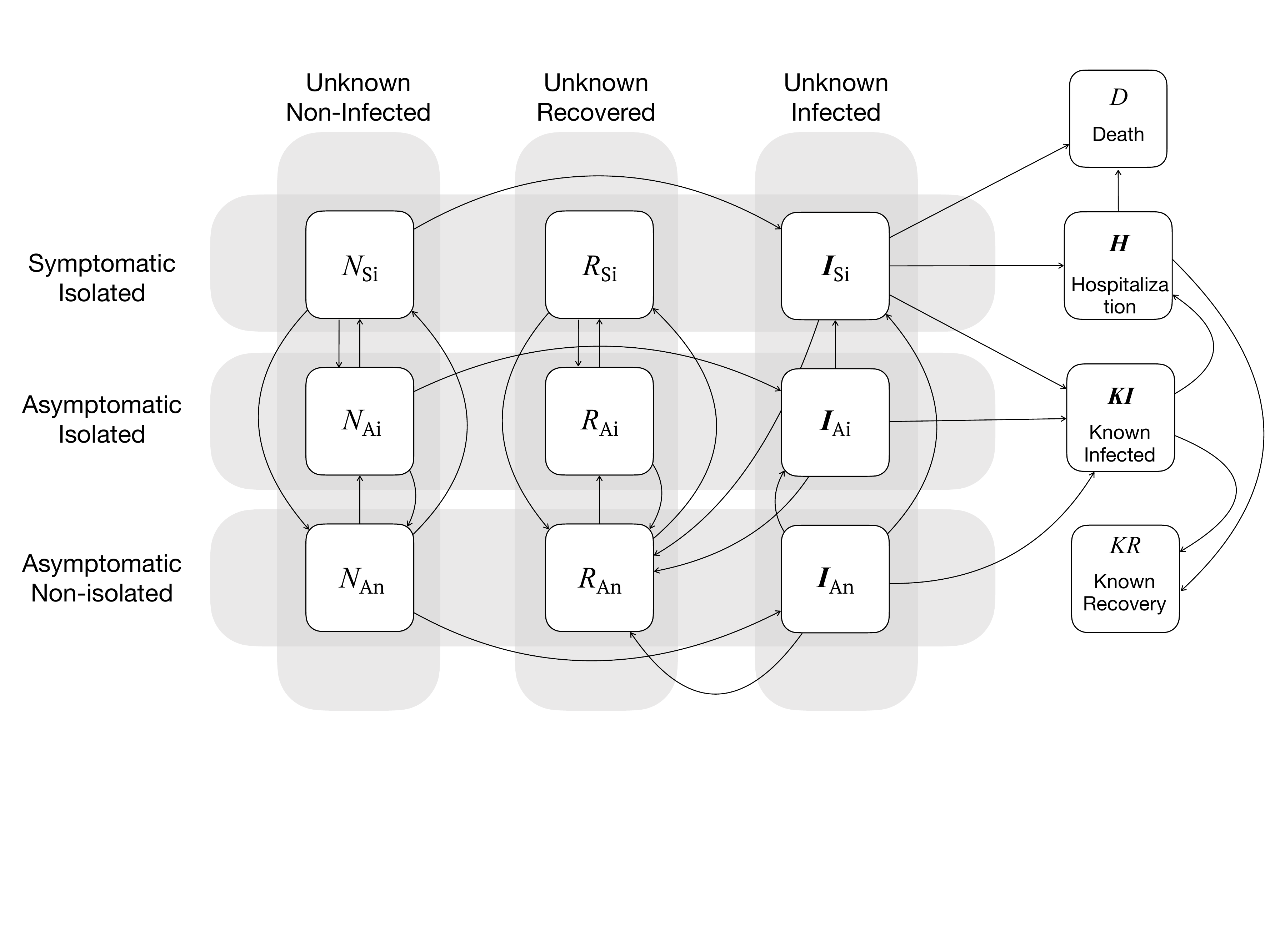}
\vspace{-5mm}
\caption{Compartments used in our model.}

\label{fig:comp}
\end{center}
\end{figure}

Our model follows the standard susceptible-infected-recovered modeling paradigm. In Figure~\ref{fig:comp}, we show the compartments in our model. The boxes represent the compartments. The arrows represent the possible flows between the compartments. The three compartments on the right of the figure capture the  individuals who know that they are infected. In particular, the compartment labeled ``KI'' corresponds to the infected individuals who know that they are infected and not (yet) hospitalized. The compartment labeled ``H'' captures the infected individuals who are hospitalized. The individuals in the latter compartment also know that they are infected. The compartment labeled ``KR'' captures the recovered individuals who knew that they were infected.~Perhaps optimistically, in our model, the individuals in all of these three compartments minimally infect others, since they are aware that they were infected.

The nine compartments on the left side of the figure capture the individuals who do not know whether they are infected. Some of these individuals will, in actuality, be infected, some will be noninfected, and some will even have recovered without knowing they had been infected.~We classify the individuals who do not know whether they are infected along two dimensions. In the first dimension, any such individual could be symptomatic-isolated, asymptomatic-isolated, or asymptomatic-nonisolated. Asymptomatic-nonisolated individuals do not show symptoms of the disease and do not make an effort to reduce their contact. Asymptomatic-isolated individuals make a conscious effort to reduce their contact, mainly because they have been in touch with a person who tested positive.~Symptomatic-isolated individuals show symptoms but do not know whether they are infected. Their symptoms may be due to other diseases. 

Once again, perhaps optimistically, in our model, all symptomatic individuals isolate themselves, so we do not have a compartment for symptomatic-nonisolated people. Symptomatic-isolated and asymptomatic-isolated individuals infect others with a smaller rate, when compared with asymptomatic-nonisolated people. When stricter social distancing norms are enforced, all individuals will reduce their contact with the general population, but we follow the assumption that asymptomatic-nonisolated individuals will still maintain a higher contact rate with the rest of the population than asymptomatic-isolated and symptomatic-isolated individuals.

Considering the individuals who do not know whether they are infected, we classify them along another dimension. In this second dimension, an individual who does not know whether she is infected could be infected, noninfected, or recovered. Thus, considering the nine compartments on the left side of Figure \ref{fig:comp}, the compartment labeled ``$I_{\text{An}}$,'' for example, captures the  infected and asymptomatic-nonisolated individuals who do not know whether they are infected. Note that these individuals are infected in actuality, but they do not know that they are infected and they maintain a high contract rate with the rest of the population, being nonisolated. When we perform tests on a segment of a population, the proportion of positive tests is given by the fraction of the infected individuals in the segment relative to the size of the whole segment. For example, using the label of a compartment to also denote the number of individuals in the compartment, since there are $N_{\text{An}} + R_{\text{An}} + I_{\text{An}}$ asymptomatic-nonisolated individuals, if we test $T$  asymptomatic-nonisolated individuals, then the number of positive tests is $
T \times \frac{I_{\text{An}}}{N_{\text{An}} + R_{\text{An}} + I_{\text{An}}}.
$

So far, we indicated two reasons for our model to be optimistic. In addition, our model assumes that close contacts of every positive case are traced and these close contacts remain in isolation for 14 days or until they are tested negative. In other words, our model implicitly assumes that the bottleneck is the testing capacity, not contact tracing capacity. It is important to note that growing the testing capacity in the City beyond a certain level without growing the contact tracing capacity will not be of much benefit, so any decision regarding how much to grow the contact tracing capacity should be directly informed by the plans to grow the testing capacity.

\section{Mathematical Description of the System Dynamics}
\label{section:system-dynamics}

\vspace{10pt}

In this section, we provide a detailed description of the system dynamics of our model.  The state of the system at each time period $t$ is described by a vector 
$$
    \left( {\bm{I}}_{\sf Si}^t, \bm{I}^t_{\sf Ai}, \bm{I}^t_{\sf An}, R^t_{\sf Si}, R^t_{\sf Ai}, R^t_{\sf An}, N^t_{\sf Si}, N^t_{\sf Ai}, N^t_{\sf An}, D^t, \bm{H}^t, \bm{KI}^t, {KR}^t  \right)~,
$$
where the variables $\bm{I}_{\sf Si}^t, \bm{I}^t_{\sf Ai}, \bm{I}^t_{\sf An}, R^t_{\sf Si}, R^t_{\sf Ai}, R^t_{\sf An}, N^t_{\sf Si}, N^t_{\sf Ai}, N^t_{\sf An}$ denote the states associated with the nine compartments on the left side of Figure \ref{fig:comp}. These nine compartments correspond to individuals whose COVID-19 status is unknown.  
The variable $D^t$ denotes the number of individuals who die in period $t$, and the variable $\bm{H}^t$ denotes the number of individuals who are hospitalized in period $t$. The variable $\bm{KI}^t$ captures the number of individuals who have been confirmed to have COVID-19 and are currently infected in period $t$. The variable ${KR}^t$ captures the number of individuals who have been confirmed to have COVID-19 and recovered by period $t$. 

{\bf Description of Sub-compartments:} For our state variables, we use regular font to denote scalars and bold font to denote vectors. We use vector notation when a compartment has multiple sub-compartments representing different subgroups of the population. Here are the descriptions of the sub-compartments.

\begin{itemize}[leftmargin=*]
    \item {Sub-compartments within the $\bm{I}_{\sf Si}$ compartment:}  The $\bm{I}_{\sf Si}$ compartment has 3 sub-compartments: (a) those who will recover naturally without requiring any hospitalization, (b) those who will require hospitalization, and (c) those who will die without access to a COVID-19 diagnostic test. Thus, 
    $$
       \bm{I}^t_{\sf Si} = \left( I^t_{\sf Si}(\mathrm{recovered}), I^t_{\sf Si}(\mathrm{hospitalized}), I^t_{\sf Si}(\mathrm{death}) \right)~.
    $$
    We use the notation  $\bar{I}^t_{\sf Si} = I^t_{\sf Si}(\mathrm{recovered}) \,+\, I^t_{\sf Si}(\mathrm{hospitalized}) \,+\, I^t_{\sf Si}(\mathrm{death})$ to denote the total number of individuals in the $\bm{I}_{\sf Si}$ compartment. The trick we use here is that when an individual is infected, we immediately decide whether this person will recover, will be hospitalized, or will die. We keep the identity of the individual accordingly throughout the simulation.

    \item {Sub-compartments within the $\bm{I}_{\sf Ai}$ compartment:}  The $\bm{I}_{\sf Ai}$ compartment has 2 sub-compartments: (a) those who will never develop COVID-19 symptoms, and (b) those who will show COVID-19 symptoms but are currently pre-symptomatic. Thus,
    $$
        \bm{I}^t_{\sf Ai} = \left( I^t_{\sf Ai}(\mathrm{recovered}), I^t_{\sf Ai}(\mathrm{show ~symptom}) \right)~,
    $$
    and we let $\bar{I}^t_{\sf Ai} ~=~ I^t_{\sf Ai}(\mathrm{recovered}) ~+~ I^t_{\sf Ai}(\mathrm{show ~symptom})$ denote the total number of individuals in the $\bm{I}_{\sf Ai}$ compartment.
    
     \item {Sub-compartments within the $\bm{I}_{\sf An}$ compartments:}  Similar to the $\bm{I}_{\sf Ai}$ compartment,  the $\bm{I}_{\sf An}$ compartment has 2 sub-compartments: (a) those who will never develop COVID-19 symptoms, and (b) those who will show COVID-19 symptoms but are currently pre-symptomatic. Thus,
    $$
         \bm{I}^t_{\sf An} = \left( I^t_{\sf An}(\mathrm{recovered}), I^t_{\sf An}(\mathrm{show ~symptom}) \right)~,
    $$
    and as before, we let $\bar{I}^t_{\sf An} ~=~ I^t_{\sf An}(\mathrm{recovered}) ~+~ I^t_{\sf An}(\mathrm{show ~symptom})$ denote the total number of individuals in the $\bm{I}_{\sf An}$ compartment.

    \item {Sub-compartments within the $\bm{H}$ compartment:}  The $\bm{H}$ compartment consists of two types of individuals: (a) those who will die and (b) those who will eventually recover. Thus, we have
    $\bm{H}^t = \left( H^t(\mathrm{die}), H^t(\mathrm{recovered}) \right)$.
    
    \item {Sub-compartments within the $\bm{KI}$ compartment:}  The $\bm{KI}$ compartment consists of two types of individuals: (a) those who will require hospitalization and (b) those who will recover without visiting a hospital. Therefore,  we have that
    $\bm{KI}^t = \left( {KI}^t(\mathrm{hospitalized}), {KI}^t(\mathrm{recovered}) \right)$.
    
\end{itemize}

{\bf Impact of Testing:} 
We assume that the diagnostic test is 100\% accurate and the result is obtained  instantaneously. However when we perform the test, we cannot differentiate between unknown infected, unknown non-infected, and recovered people.  We can only test people based on their observable characteristics.  At the beginning of each period, we test $T_{\sf Si}$ symptomatic-isolated individuals, $T_{\sf Ai}$ asymptomatic-isolated individuals, and $T_{\sf An}$ asymptomatic-nonisolated individuals.  We assume that the tests are administered at random within each of these populations. Let $\pi_{\sf Si}$, $\pi_{\sf Ai}$, and $\pi_{\sf An}$ denote the fraction of symptomatic isolated, asymptomatic isolated, and  asymptomatic non-isolated individuals, respectively, who receive diagnostic tests.  Then,
\begin{align*}
     \pi_{\sf Si}  &=  \min \left\{ \frac{ T^t_{\sf Si} }{ \displaystyle \bar{I}^t_{\sf Si} ~-~ I^t_{{\sf Si}}(\mathrm{death}) ~+~ R^t_{\sf Si} ~+~ N^t_{\sf Si} } ~,~ 1 \right\},\\
     \pi_{\sf Ai}  &=  \min \left\{ \frac{ T^t_{\sf Ai} }{ \displaystyle \bar{I}^t_{\sf Ai} ~+~ R^t_{\sf Ai} + N^t_{\sf Ai} } ~,~ 1 \right\}, \\
     \pi_{\sf An}  &=  \min \left\{ \frac{ T^t_{\sf An} }{ \displaystyle \bar{I}^t_{\sf An}~+~ R^t_{\sf An} + N^t_{\sf An} }  ~,~ 1 \right\}~.
\end{align*}
The above  fractions drive the dynamics across compartments, which are described below based on the events that can occur in our~model. In the fraction $\pi_{\sf Si}$, we assume that the infected symptomatic-isolated individuals from the death sub-compartment are inaccessible for testing.

{\bf Description of Dynamics Between Compartments:} Our description below is organized by the compartments.  For each compartment, we describe the inflows, outflows, and the new state at time period $t + 1$.  

\begin{itemize}[leftmargin=*]

\item {\bf Infected Symptomatic-Isolated Compartment ($\bm{I}_{\sf Si}$):}  As noted earlier,  the $\bm{I}_{\sf Si}$ compartment has 3 subgroups of individuals: (a) those who will recover naturally without requiring any hospitalization, (b) those who require hospitalization, and (c) those who die without access to a COVID-19 diagnostic test. Thus,
 $$
       \bm{I}^t_{\sf Si} = \left( I^t_{\sf Si}(\mathrm{recovered}), I^t_{\sf Si}(\mathrm{hospitalized}), I^t_{\sf Si}(\mathrm{death}) \right)~,
    $$
and   $\bar{I}^t_{\sf Si} = I^t_{\sf Si}(\mathrm{recovered}) \,+\, I^t_{\sf Si}(\mathrm{hospitalized}) \,+\, I^t_{\sf Si}(\mathrm{death})$ denotes the total number of individuals across the three sub-compartments.


\uline{Inflow}: The inflow to the ${\bm I}_{\sf Si}$ compartment consists of 3 sources given by:
\begin{itemize}[leftmargin=*]
    \item Non-infected symptomatic-isolated ($N_{\sf Si})$ individuals who did not get tested and were newly infected during period $t$. The number of such individuals is given by
\begin{align*}
  \beta_\ell \times \Big(  \bar{I}^t_{\sf An} (1-\pi_{\sf An} )
 + \bar{I}^t_{\sf Ai} \, (1-\pi_{\sf Ai} )  +  \bar{I}^t_{\sf Si} \, (1-\pi_{\sf Si} ) \Big) \times  N^t_{\sf Si}\,(1-\,\pi_{\sf Si} )   ~,
\end{align*}
where $ \bar{I}^t_{\sf An} (1-\pi_{\sf An} )
 + \bar{I}^t_{\sf Ai} \, (1-\pi_{\sf Ai} )  +  \bar{I}^t_{\sf Si} \, (1-\pi_{\sf Si} ) $ represents the infected population who remains untested, and $\beta_\ell$ represents the effective contact rate among isolated individuals. At the start of our simulation, we set $\beta_\ell = 1.8823 \times 10^{-8}$, and starting on March 22, we reduce $\beta_\ell$ by a third, to a new value of $6.2742 \times 10^{-9}$; see Section \ref{section:model-parameters} for more details.
 
    \item Infected asymptomatic-nonisolated ($\bm{I}_{\sf An}$) individuals who become symptomatic. The total inflow of such individuals is given by 
$$
    \frac{ I^t_{\sf An}(\mathrm{show~symptom}) }{  {\tt infToSympTime}}~,
$$
    where {\tt infToSympTime} represents the average number of days until an infected person displays symptoms. In our model, we set ${\tt infToSympTime}$ to be $5$ days; see Section \ref{section:model-parameters} for more details.
    
    \item Infected asymptomatic-isolated ($\bm{I}_{\sf Ai}$) individuals who become symptomatic. Using a similar logic as above, the total amount of such inflow is given by $\displaystyle \frac{I^t_{\sf Ai}(\mathrm{show~symptom})}{  {\tt infToSympTime}}$.
\end{itemize}

We assume that a fraction {\tt hospOutOfSympFrac}, currently set at 20\%, of the total inflow to the $\bm{I}_{\sf Si}$ compartment will require hospitalization, and a fraction {\tt deathOutOfSympFrac} (2\%) will die without being tested.  The remaining $1 - {\tt hospOutOfSympFrac} -  {\tt deathOutOfSympFrac} = 78\%$ will recover at home. Justifications for these fractions are given in Section \ref{section:model-parameters}.

\vspace{10pt}
\uline{Outflow}: There are four destinations for the outflow from   the $\bm{I}_{\sf Si}$ compartment.

\begin{itemize}[leftmargin=*]
\item Individuals  in the ``recovered'' sub-compartment, $I^t_{\sf Si}(\mathrm{recovered})$, will move to the recovered asymptomatic-nonisolated ($R_{\sf An}$) compartment at the   rate of $1/{\tt sympToRecoveryTime}$, where {\tt sympToRecoveryTime} represents the average time for a symptomatic individual to recover from the disease. Currently, {\tt sympToRecoveryTime} is  set at 14 days.

\item Individuals in the ``hospitalized'' sub-compartment, $I^t_{\sf Si}(\mathrm{hospitalized})$, will move to the $\bm{H}$ compartment at the rate of $1/{\tt sympToHospTime}$, where ${\tt sympToHospTime}$ (5 days)  represents the average number of days from symptom onset until hospitalization.

\item Individuals in the  ``death'' sub-compartment, $I^t_{\sf Si}(\mathrm{death})$, will move to the death compartment ($D$) at the rate of $1/{\tt sympToDeathTime}$, where {\tt sympToDeathTime} (14 days) denotes the average time from symptoms onset to death.

\item Finally, a fraction $\pi_{\sf Si}$ in every sub-compartment will move to the known infected ($\bm{KI}$)~compartment due to testing.
\end{itemize}

\vspace{10pt}
\uline{Update Equations}: 
Let ${\tt recoverOutOfSympFrac} = 1 - {\tt hospOutOfSympFrac} -  {\tt deathOutOfSympFrac}$.  We have the following update equations.

{\fontsize{10}{10}\selectfont{
\begin{align*}
&I^{t+1}_{\sf Si}(\mathrm{recovered}) \\
&~=~ I^{t}_{\sf Si}(\mathrm{recovered}) \times ( 1 - \pi_{\sf Si}  )   \times \left( 1 - \frac{1}{{\tt sympToRecoveryTime}} \right) \\
&\quad~~~~+~  {\tt recoverOutOfSympFrac}  \times \Bigg[ \beta_\ell \times
 \Big(  \bar{I}^t_{\sf An} (1-\pi_{\sf An} )
 + \bar{I}^t_{\sf Ai} \, (1-\pi_{\sf Ai} )  +  \bar{I}^t_{\sf Si} \, (1-\pi_{\sf Si} ) \Big) \times  N^t_{\sf Si}\,(1-\,\pi_{\sf Si} )  \\
&~~~~~~~~~~~~~~~~~~~~~~~~~~~~~~~~~~~~~~~~~~~~~~~~~ ~+~  \frac{I^t_{\sf An}(\mathrm{show~symptom}) (1-\pi_{\sf An})}{{\tt infToSympTime}}   ~+~ \frac{I^t_{\sf Ai}(\mathrm{show~symptom}) (1 - \pi_{\sf Ai})}{{\tt infToSympTime}}  \Bigg]~, \\
&I^{t+1}_{\sf Si}(\mathrm{hospitalized}) \\
&~=~  I^{t}_{\sf Si}(\mathrm{hospitalized}) \times ( 1 - \pi_{\sf Si}  )   \times \left( 1 - \frac{1}{{\tt sympToHospTime}} \right) \\
&\quad~~~~+~  {\tt hospOutOfSympFrac}  \times \Bigg[  \beta_\ell \times
 \Big(  \bar{I}^t_{\sf An} (1-\pi_{\sf An} )
 + \bar{I}^t_{\sf Ai} \, (1-\pi_{\sf Ai} )  +  \bar{I}^t_{\sf Si} \, (1-\pi_{\sf Si} ) \Big) \times  N^t_{\sf Si}\,(1-\,\pi_{\sf Si} )  \\
&~~~~~~~~~~~~~~~~~~~~~~~~~~~~~~~~~~~~~~~~~~~~~~ ~+~  \frac{I^t_{\sf An}(\mathrm{show~symptom}) (1-\pi_{\sf An})}{{\tt infToSympTime}}   ~+~ \frac{I^t_{\sf Ai}(\mathrm{show~symptom}) (1 - \pi_{\sf Ai})}{{\tt infToSympTime}}  \Bigg]~,\\
&I^{t+1}_{\sf Si}(\mathrm{death}) \\
&~=~  I^{t}_{\sf Si}(\mathrm{death}) \times ( 1 - \pi_{\sf Si}  )   \times \left( 1 - \frac{1}{{\tt sympToDeathTime}} \right) \\
&\quad~~~~+~ {\tt deathOutOfSympFrac} \times \Bigg[  \beta_\ell \times
 \Big(  \bar{I}^t_{\sf An} (1-\pi_{\sf An} )
 + \bar{I}^t_{\sf Ai} \, (1-\pi_{\sf Ai} )  +  \bar{I}^t_{\sf Si} \, (1-\pi_{\sf Si} ) \Big) \times N^t_{\sf Si}\,(1-\,\pi_{\sf Si} )  \\
&~~~~~~~~~~~~~~~~~~~~~~~~~~~~~~~~~~~~~~~~~~~~~~ ~+~  \frac{I^t_{\sf An}(\mathrm{show~symptom}) (1-\pi_{\sf An})}{{\tt infToSympTime}}   ~+~ \frac{I^t_{\sf Ai}(\mathrm{show~symptom}) (1 - \pi_{\sf Ai})}{{\tt infToSympTime}}  \Bigg]~.
\end{align*}
}}

\vspace{10pt}
\item{{\bf Infected Asymptomatic-Isolated Compartment ($\bm{I}_{\sf Ai}$)}:}
We have two sub-compartments with $\bm{I}^t_{\sf Ai} = \left( I^t_{\sf Ai}(\mathrm{recovered}), I^t_{\sf Ai}(\mathrm{show ~symptom}) \right)$, and \mbox{$\bar{I}^t_{\sf Ai} = I^t_{\sf Ai}(\mathrm{recovered}) ~+~ I^t_{\sf Ai}(\mathrm{show ~symptom})$} denotes the total number of individuals across the two sub-compartments.

\vspace{10pt}
\uline{Inflow}: The inflow to the $\bm{I}_{\sf Ai}$ compartment has two sources:
\begin{itemize}[leftmargin=*]
    \item    Non-infected asymptomatic-isolated ($N_{\sf Ai}$) individuals who were not tested and became newly infected during period $t$. The number of such individuals is given by
\begin{align*}
 \beta_\ell \times \Big( \bar{I}^t_{\sf An} (1-\pi_{\sf An} )
 + \bar{I}^t_{\sf Ai} \, (1-\pi_{\sf Ai} )  + \bar{I}^t_{\sf Si} \, (1-\pi_{\sf Si} ) \Big) \times   N^t_{\sf Ai}\,(1-\,\pi_{\sf Ai} )   ~,
\end{align*}
where $\beta_\ell$ denotes the effective contact rate for isolated individuals.

    \item  Infected asymptomatic-nonisolated ($\bm{I}_{\sf An}$) individuals who were not tested but were identified as a close contact of a positive case. These individuals voluntarily self isolate and reduce their contact rate with others. The number of such individuals is given by
 $$
     {\tt contactPerPosCase} \times  \frac{  \bar{I}^t_{\sf Si} \pi_{\sf Si} + \bar{I}^t_{\sf Ai} \pi_{\sf Ai}  + \bar{I}^t_{\sf An} \pi_{\sf An} }{\bar{H}} \times  {\tt likOfBeingInfected} \times  \bar{I}^t_{\sf An}(1-\pi_{\sf An})~,
 $$
where {\tt contactPerPosCase} is the average number of contacts per positive case, currently set at $4$, whereas $\bar{I}^t_{\sf Si} \pi_{\sf Si} + \bar{I}^t_{\sf Ai} \pi_{\sf Ai}  + \bar{I}^t_{\sf An} \pi_{\sf An}$ denotes the number of positive cases identified that time period and $\bar{H}$ is given by
{\fontsize{10}{10}\selectfont{
$$
\bar{H} = \left( {\tt likOfBeingInfected} \times \bar{I}^t_{\sf An}(1-\pi_{\sf An})\right) \,+\, (R^t_{\sf An} ~+~ R^t_{\sf Si} \pi_{\sf Si} ~+~ R^t_{\sf Ai}\pi_{\sf Ai}) \,+\, (N^t_{\sf An} ~+~ N^t_{\sf Si} \pi_{\sf Si} ~+~ N^t_{\sf Ai}\pi_{\sf Ai})~,
$$}}
\!\!\!where the parameter {\tt likOfBeingInfected} measures the likelihood that a contacted individual is in the infected population as compared to the non-infected and recovered populations. We set {\tt likOfBeingInfected} at 5.5.
\end{itemize}
We assume that a fraction {\tt symptomFrac}, currently set at 50\%, of  the total inflow to the $\bm{I}_{\sf Ai}$ compartment will develop symptoms, and the remaining $1 - {\tt symptomFrac} = 50\%$ will not develop any symptoms. 

\vspace{10pt}
\uline{Outflow}: There are three destinations for the outflow from the $\bm{I}_{\sf Ai}$ compartment.
\begin{itemize}[leftmargin=*]
    \item The individuals  in the ``recovered'' sub-compartment, $I^t_{\sf Ai}(\mathrm{recovered})$, will move to the $R_{\sf An}$ compartment at the rate of $1/{\tt asympToRecoveryTime}$. The parameter {\tt asympToRecoveryTime} represents the average time that an asymptomatic infected person self-isolates after being identified as a close contact. We currently set this value to 10 days.
    
    \item The ``show symptom'' sub-compartment, $I^t_{\sf Ai}(\mathrm{show~symptom})$, will move to the $\bm{I}_{\sf Si}$ compartment at the rate of $1/{\tt infToSympTime}$, where {\tt infToSympTime} (5 days) is the average time until symptom onset.

    \item A fraction $\pi_{\sf Ai}$ of individuals who are tested will move to the known infected ($\bm{KI}$)~compartment.
\end{itemize}

\vspace{10pt}
\uline{Update Equations}: We have the following update equations.
{\fontsize{10}{10}\selectfont{
\begin{align*}
&I^{t+1}_{\sf Ai}(\mathrm{recovered}) \\
&~=~ 
I^{t}_{\sf Ai}(\mathrm{recovered})\times ( 1 - \pi_{\sf Ai}  )   \times \left( 1 - \frac{1}{{\tt asympToRecoveryTime}} \right)
\\
&\quad~~~~~+~  \left( 1 - {\tt symptomFrac} \right)  \times \Bigg[  \beta_\ell \times \Big( \bar{I}^t_{\sf An} (1-\pi_{\sf An} )
 + \bar{I}^t_{\sf Ai} \, (1-\pi_{\sf Ai} )  + \bar{I}^t_{\sf Si} \, (1-\pi_{\sf Si} ) \Big) \times   N^t_{\sf Ai}\,(1-\,\pi_{\sf Ai} )     \\
&~~~~~~~~~~~~~~~ ~+~       {\tt contactPerPosCase} \times  \frac{  \bar{I}^t_{\sf Si} \pi_{\sf Si} + \bar{I}^t_{\sf Ai} \pi_{\sf Ai}  + \bar{I}^t_{\sf An} \pi_{\sf An} }{\bar{H}} \times  {\tt likOfBeingInfected} \times  \bar{I}^t_{\sf An}(1-\pi_{\sf An}) \bigg]~, \\
&I^{t+1}_{\sf Ai}(\mathrm{show~symptom}) \\
&~=~  I^{t}_{\sf Ai}(\mathrm{show~symptom}) \times ( 1 - \pi_{\sf Ai}  )   \times \left( 1 - \frac{1}{{\tt infToSympTime}} \right) \\
&\quad~~~~~+~  {\tt symptomFrac}   \times \Bigg[  \beta_\ell \times  \Big( \bar{I}^t_{\sf An} (1-\pi_{\sf An} )
 + \bar{I}^t_{\sf Ai} \, (1-\pi_{\sf Ai} )  + \bar{I}^t_{\sf Si} \, (1-\pi_{\sf Si} ) \Big) \times  N^t_{\sf Ai}\,(1-\,\pi_{\sf Ai} )     \\
&~~~~~~~~~~~~~~~ ~+~       {\tt contactPerPosCase} \times  \frac{  \bar{I}^t_{\sf Si} \pi_{\sf Si} + \bar{I}^t_{\sf Ai} \pi_{\sf Ai}  + \bar{I}^t_{\sf An} \pi_{\sf An} }{\bar{H}} \times  {\tt likOfBeingInfected} \times  \bar{I}^t_{\sf An}(1-\pi_{\sf An}) \bigg]~.
\end{align*}
}}


\vspace{10pt}
\item{{\bf Infected Asymptomatic-Nonisolated Compartment ($\bm{I}_{\sf An}$)}:}
We have two sub-compartments, $\bm{I}^t_{\sf An} = \left( I^t_{\sf An}(\mathrm{recovered}), I^t_{\sf An}(\mathrm{show ~symptom}) \right)$, and we denote the total number of individuals across the two sub-compartments as $\bar{I}^t_{\sf An} = I^t_{\sf An}(\mathrm{recovered}) ~+~ I^t_{\sf An}(\mathrm{show ~symptom})$.  

\vspace{10pt}
\uline{Inflow}: The inflow to the $\bm{I}_{\sf An}$ compartment comes from non-infected asymptomatic-nonisolated $N_{\sf An}$ individuals who become newly infected during period $t$. The number of such individuals is given by
\begin{align*}
 \Big( \beta_h \bar{I}^t_{\sf An} (1-\pi_{\sf An} )
 + \beta_\ell \bar{I}^t_{\sf Ai} \, (1-\pi_{\sf Ai} )  + \beta_\ell \bar{I}^t_{\sf Si} \, (1-\pi_{\sf Si} ) \Big) \times N^t_{\sf An}\,(1-\,\pi_{\sf An} )   ~,
\end{align*}
where $\beta_h$ represents the effective contact rate among nonisolated individuals. At the start of our simulation, we set $\beta_h = 2.82341 \times 10^{-8}$, and starting on March 22, we reduce $\beta_h$ by a third, to a new value of $9.41135 \times 10^{-9}$; see Section \ref{section:model-parameters} for more details. We assume that a fraction {\tt symptomFrac}, currently set at 50\%, of  the total inflow to the $\bm{I}_{\sf An}$ compartment will develop symptom, and the remaining $1 - {\tt symptomFrac} = 50\%$ will not develop any symptoms. 


\vspace{10pt}
\uline{Outflow}: The outflow from the $\bm{I}_{\sf An}$ compartment has four destinations.
\begin{itemize}[leftmargin=*]
    \item The individuals  in the ``recovered'' sub-compartment, $I^t_{\sf An}(\mathrm{recovered})$, will move to the $R_{\sf An}$ compartment at the rate of $1/{\tt asympToRecoveryTime}$.
    
    \item The individuals  in the ``show symptom'' sub-compartment, $I^t_{\sf An}(\mathrm{show~symptom})$, will move to the $\bm{I}_{\sf Si}$ compartment at the rate of $1/{\tt infToSympTime}$.
    
    \item   Some individuals who were not tested will reduce their contacts and move to the $\bm{I}_{\sf Ai}$ compartment. The number of such individuals is
$$
     {\tt contactPerPosCase} \times  \frac{  \bar{I}^t_{\sf Si} \pi_{\sf Si} + \bar{I}^t_{\sf Ai} \pi_{\sf Ai}  + \bar{I}^t_{\sf An} \pi_{\sf An} }{\bar{H}} \times  {\tt likOfBeingInfected} \times  \bar{I}^t_{\sf An}(1-\pi_{\sf An})~.
 $$

    \item A fraction $\pi_{\sf An}$ of individuals are tested and moved to the known infected $\bm{KI}$ compartment.
\end{itemize}

\vspace{10pt}
\uline{Update Equations}: We have the following update equations.
{\fontsize{10}{10}\selectfont{
\begin{align*}
&I^{t+1}_{\sf An}(\mathrm{recovered}) \\
&~=~ I^{t}_{\sf An}(\mathrm{recovered}) \times ( 1 - \pi_{\sf An})   \times  \bigg( 1 - \frac{1}{{\tt asympToRecoveryTime}}  \\
&~~~~~~~~~~~~~~~~~~~~~~~~~~~~~~~ ~-~   {\tt contactPerPosCase} \times  \frac{  \bar{I}^t_{\sf Si} \pi_{\sf Si} + \bar{I}^t_{\sf Ai} \pi_{\sf Ai}  + \bar{I}^t_{\sf An} \pi_{\sf An} }{\bar{H}} \times  {\tt likOfBeingInfected} \bigg)\\
&\quad~~~~+~  (1 ~-~ {\tt symptomFrac}) \times \bigg[   \Big( \beta_h \bar{I}^t_{\sf An} (1-\pi_{\sf An} )
 + \beta_\ell \bar{I}^t_{\sf Ai} \, (1-\pi_{\sf Ai} )  + \beta_\ell \bar{I}^t_{\sf Si} \, (1-\pi_{\sf Si} ) \Big) \times N^t_{\sf An}\,(1-\,\pi_{\sf An} )   \bigg]~, \\
&I^{t+1}_{\sf An}(\mathrm{show~symptom}) \\
&~=~  I^{t}_{\sf An}(\mathrm{show~symptom}) \times ( 1 - \pi_{\sf An}  )   \times \bigg( 1 - \frac{1}{{\tt infToSympTime}} \\
&~~~~~~~~~~~~~~~~~~~~~~~~~~~~~~~ ~-~   {\tt contactPerPosCase} \times  \frac{  \bar{I}^t_{\sf Si} \pi_{\sf Si} + \bar{I}^t_{\sf Ai} \pi_{\sf Ai}  + \bar{I}^t_{\sf An} \pi_{\sf An} }{\bar{H}} \times  {\tt likOfBeingInfected} \bigg)\\
&\quad~~~~+~  {\tt symptomFrac} \times \bigg[   \Big( \beta_h \bar{I}^t_{\sf An} (1-\pi_{\sf An} )
 + \beta_\ell \bar{I}^t_{\sf Ai} \, (1-\pi_{\sf Ai} )  + \beta_\ell \bar{I}^t_{\sf Si} \, (1-\pi_{\sf Si} ) \Big) \times N^t_{\sf An}\,(1-\,\pi_{\sf An} )   \bigg]~.
\end{align*}
}}


\vspace{10pt}
\item{{\bf Non-infected Symptomatic-Isolated Compartment ($N_{\sf Si}$)}:} 

\vspace{10pt}
\uline{Inflow}: The inflow to the $N_{\sf Si}$ compartment has two sources:
\begin{itemize}[leftmargin=*]
    \item Non-infected asymptomatic-nonisolated ($N_{\sf An}$) individuals who did not get tested and developed symptoms caused by another condition or disease such as the seasonal flu. There are $N^t_{\sf An} ~+~ N^t_{\sf Si} \pi_{\sf Si} ~+~ N^t_{\sf Ai}\pi_{\sf Ai}$ such individuals. The rate that an individual develops symptoms due to a non-COVID-19 disease is given by the parameter {\tt nonCOVIDSymptRate}, which we currently set at 1/1200; see Section \ref{section:model-parameters} for more details on how we arrive at this number.
    
    \item Non-infected asymptomatic-isolated ($N_{\sf Ai})$ individuals can also develop flu-like symptoms. There are $N^{t}_{\sf Ai} \times ( 1 - \pi_{\sf Ai})$ such individuals.
\end{itemize}

\vspace{10pt}
\uline{Outflow}: The outflow from the $N_{\sf Si}$ compartment has three destinations:
\begin{itemize}[leftmargin=*]
    \item Individuals  who are tested will return negative and move to the $N_{\sf An}$ compartment. This causes them to increase their contact with other individuals.
    \item Some individuals will become infected from coming in contact with infected people.
    \item Individuals leave self-quarantine at a rate of $1/{\tt selfQuarTime}$, where {\tt selfQuarTime} is the average time that an individual self-isolates. We currently set this value to 10 days.
\end{itemize}

\vspace{10pt}
\uline{Update Equations}: We have the following update equations.
{\fontsize{10}{10}\selectfont{
\begin{align*}
N^{t+1}_{\sf Si} &~=~ N^{t}_{\sf Si} \times ( 1 - \pi_{\sf Si})   \times \left( 1 ~-~ \dfrac{1}{{\tt selfQuarTime}} ~-~ \beta_\ell \times \Big(  \bar{I}^t_{\sf An} (1-\pi_{\sf An} )
 + \bar{I}^t_{\sf Ai} \, (1-\pi_{\sf Ai} )  +  \bar{I}^t_{\sf Si} \, (1-\pi_{\sf Si} ) \Big)  \right) \\
 &\quad~~~~+~  {\tt nonCOVIDSymptRate} \times \left(  N^t_{\sf Si} \pi_{\sf Si} ~+~ N^t_{\sf Ai} ~+~ N^t_{\sf An} \right)~. 
\end{align*}
}}

\vspace{10pt}
\item{{\bf Non-infected Asymptomatic-Isolated Compartment ($N_{\sf Ai}$)}:}

\vspace{10pt}
\uline{Inflow}: The inflow to the $N_{\sf Ai}$ compartment comes from individuals in the $N_{\sf An}$ compartment who reduce their exposure to other individuals after being identified as a close contact. The number of such individuals is given~by
$$
    {\tt contactPerPosCase} \times  \frac{  \bar{I}^t_{\sf Si} \pi_{\sf Si} + \bar{I}^t_{\sf Ai} \pi_{\sf Ai}  + \bar{I}^t_{\sf An} \pi_{\sf An} }{\bar{H}} \times  \left( N^t_{\sf An} ~+~ N^t_{\sf Si} \pi_{\sf Si} ~+~ N^t_{\sf Ai} \pi_{\sf Ai} \right)~,
$$
where {\tt contactPerPosCase} is the average number of contacts per positive case, currently set at~$4$, and  $\bar{H}$ denotes the total number of high-contact individuals; see the discussion in the $\bm{I}_{\sf Ai}$ compartment.

\vspace{10pt}
\uline{Outflow}: The outflow from the $N_{\sf Ai}$ compartment has four destinations.
\begin{itemize}[leftmargin=*]
\item Individuals who are tested will test negative and move to the $N_{\sf An}$ compartment, end their self-isolation, and increase their contacts.
\item Untested individuals leave self-isolation at a rate of $1/{\tt selfQuarTime}$.
\item Some individuals will become infected and move to the $\bm{I}_{\sf Ai}$ compartment.  
\item Some individuals will develop flu-like symptoms but not COVID-19, at the rate of  {\tt nonCOVIDSymptRate}.
\end{itemize}

\vspace{10pt}
\uline{Update Equations}: We have the following update equations.
\begin{align*}
N^{t+1}_{\sf Ai} &~=~ N^{t}_{\sf Ai} \times ( 1 - \pi_{\sf Ai})   \times \bigg( 1 ~-~ \frac{1}{{\tt selfQuarTime}} ~-~ {\tt nonCOVIDSymptRate} \\
&~~~~~~~~~~~~~~~~~~~~~~~~~~~~~~~~~ ~-~  \beta_\ell \times \Big( \bar{I}^t_{\sf An} (1-\pi_{\sf An} )
 + \bar{I}^t_{\sf Ai} \, (1-\pi_{\sf Ai} )  + \bar{I}^t_{\sf Si} \, (1-\pi_{\sf Si} ) \Big)  \bigg) \\
 &\quad~~~~~+~       {\tt contactPerPosCase} \times  \frac{  \bar{I}^t_{\sf Si} \pi_{\sf Si} + \bar{I}^t_{\sf Ai} \pi_{\sf Ai}  + \bar{I}^t_{\sf An} \pi_{\sf An} }{\bar{H}} \times  \left( N^t_{\sf An} ~+~ N^t_{\sf Si} \pi_{\sf Si} ~+~ N^t_{\sf Ai} \pi_{\sf Ai} \right)~.
\end{align*}


\vspace{10pt}
\item{{\bf Non-infected Asymptomatic-Nonisolated Compartment ($N_{\sf An}$)}:}

\vspace{10pt}
\uline{Inflow}: The inflow to the $N_{\sf An}$ compartment comes from three sources. 
First, individuals from the $N_{\sf Si}$ compartment recover from non-COVID-19 symptoms at a rate of $\dfrac1{\tt selfQuarTime}$. 
Second, individuals from the $N_{\sf Ai}$ compartment leave self-isolation and increase their contact with others at a rate of ${\dfrac1{\tt selfQuarTime}}$.   
Third, individuals from the $N_{\sf Si}$ and $N_{\sf Ai}$ compartments who test negative will also increase their contact rates.

\vspace{10pt}
\uline{Outflow}: The outflow from the $N_{\sf An}$ compartment has three destinations.  First, some individuals are identified as close contacts and move to the $N_{\sf Ai}$ compartment, thereby reducing their contact levels. Second, other individuals will develop flu-like symptoms independent of COVID and move to the $\bm{N}_{\sf Si}$ compartment at the rate of {\tt nonCOVIDSymptRate}. Third,  some individuals will become infected and move to the $\bm{I}_{\sf An}$ compartment.

\vspace{10pt}
\uline{Update Equations}: We have the following update equations.
\begin{align*}
N^{t+1}_{\sf An} &~=~ \left( N^t_{\sf An} ~+~ N^t_{\sf Si} \pi_{\sf Si} ~+~ N^t_{\sf Ai}\pi_{\sf Ai} \right)  \\ 
&~~~~~~~~~~~~ \times \Bigg( 1 ~-~ {\tt nonCOVIDSymptRate}  ~-~   {\tt contactPerPosCase} \times  \frac{  \bar{I}^t_{\sf Si} \pi_{\sf Si} + \bar{I}^t_{\sf Ai} \pi_{\sf Ai}  + \bar{I}^t_{\sf An} \pi_{\sf An} }{\bar{H}}   \\
& ~~~~~~~~~~~~~~~~~~~~ ~-~  \Big( \beta_h \bar{I}^t_{\sf An} (1-\pi_{\sf An} )
 + \beta_\ell \bar{I}^t_{\sf Ai} \, (1-\pi_{\sf Ai} )  + \beta_\ell \bar{I}^t_{\sf Si} \, (1-\pi_{\sf Si} ) \Big) 
\Bigg) \\
 &\quad~~~~~~~~~+~       \frac{N_{\sf Ai}^t (1-\pi_{\sf Ai} )}{{\tt selfQuarTime}}  ~+~   \frac{ N_{\sf Si}^t (1-\pi_{\sf Si}) }{{\tt selfQuarTime}} ~.
\end{align*}


\vspace{10pt}
\item{{\bf Recovered Symptomatic-Isolated Compartment ($R_{\sf Si}$)}:}

\vspace{10pt}
\uline{Inflow}: The inflow to the $R_{\sf Si}$ compartment has two sources. Recovered asymptomatic-nonisolated ($R_{\sf An}$) and recovered asymptomatic-isolated ($R_{\sf Ai}$) individuals, who did not get tested, can develop flu-like symptoms independent of COVID-19, at the rate of {\tt nonCOVIDSymptRate}.

\vspace{10pt}
\uline{Outflow}: The outflow from the $R_{\sf Si}$ compartment occurs from  individuals who recover from non-COVID-19 symptoms at the rate of $1/{\tt selfQuarTime}$.

\vspace{10pt}
\uline{Update Equations}: We have the following update equations.
{\fontsize{10}{10}\selectfont{
\begin{align*}
R^{t+1}_{\sf Si} &~=~ R^{t}_{\sf Si} \times ( 1 - \pi_{\sf Si})   \times \left( 1 ~-~ \frac{1}{{\tt selfQuarTime}} \right)  ~+~ {\tt nonCOVIDSymptRate} \times \left(  R^t_{\sf Si} \pi_{\sf Si} ~+~ R^t_{\sf Ai} ~+~ R^t_{\sf An} \right) ~.
\end{align*}
}}


\vspace{10pt}
\item{{\bf Recovered Asymptomatic-Isolated Compartment ($R_{\sf Ai}$)}:}

\vspace{10pt}
\uline{Inflow}: The inflow to the $R_{\sf Ai}$ compartment  comes from individuals in the $R_{\sf An}$ compartment who reduce their exposure to other individuals after being identified as a close contact.    The number of such individuals is given by
$$
    {\tt contactPerPosCase} \times  \frac{  \bar{I}^t_{\sf Si} \pi_{\sf Si} + \bar{I}^t_{\sf Ai} \pi_{\sf Ai}  + \bar{I}^t_{\sf An} \pi_{\sf An} }{\bar{H}} \times  \left( R^t_{\sf An} ~+~ R^t_{\sf Si} \pi_{\sf Si} ~+~ R^t_{\sf Ai} \pi_{\sf Ai} \right)~,
$$

\vspace{10pt}
\uline{Outflow}: The outflow from the $R_{\sf Ai}$ compartment has two destinations. First, untested individuals leave isolation at the rate of $1/{\tt selfQuarTime}$. Second, other individuals will develop symptoms independent of COVID-19, at the rate of {\tt nonCOVIDSymptRate}.

\vspace{10pt}
\uline{Update Equations}: We have the following update equations.
\begin{align*}
R^{t+1}_{\sf Ai} &~=~ R^{t}_{\sf Ai} \times ( 1 - \pi_{\sf Ai})   \times \left( 1 ~-~ \frac{1}{{\tt selfQuarTime}} ~-~  {\tt nonCOVIDSymptRate}  \right) \\
 &\quad~~~~~+~     {\tt contactPerPosCase} \times  \frac{  \bar{I}^t_{\sf Si} \pi_{\sf Si} + \bar{I}^t_{\sf Ai} \pi_{\sf Ai}  + \bar{I}^t_{\sf An} \pi_{\sf An} }{\bar{H}} \times  \left( R^t_{\sf An} ~+~ R^t_{\sf Si} \pi_{\sf Si} ~+~ R^t_{\sf Ai} \pi_{\sf Ai} \right)~.
\end{align*}


\vspace{10pt}
\item{{\bf Recovered Asymptomatic-Nonisolated Compartment ($R_{\sf An}$)}:}

\vspace{10pt}
\uline{Inflow}: The inflow to the $R_{\sf An}$ compartment has six sources:
\begin{itemize}
    \item  Untested people from the $\bm{I}_{\sf Si}$ compartment who recovered naturally.
    \item  Untested people from the $\bm{I}_{\sf Ai}$ compartment who recovered naturally.
    \item  Untested people from the $\bm{I}_{\sf An}$ compartment who recovered naturally.
        \item People from the $R_{\sf Ai}$ and $R_{\sf Si}$ compartments who test negative.
    \item  People from the $R_{\sf Ai}$ compartment who leave self-quarantine.
    \item People from the $R_{\sf Si}$ compartment whose non-COVID-19 symptoms disappeared.
\end{itemize}

\vspace{10pt}
\uline{Outflow}: The outflow from the $R_{\sf An}$ compartment occurs in two ways.  First,
some individuals are identified as close contacts and move to the $R_{\sf Ai}$ compartment. Second, some individuals will develop symptoms independent of COVID-19, at the rate of {\tt nonCOVIDSymptRate}.

\vspace{10pt}
\uline{Update Equations}: We have the following update equations.
{\fontsize{10}{10}\selectfont{
\begin{align*}
R^{t+1}_{\sf An} &~=~ \left( R^t_{\sf An} ~+~ R^t_{\sf Si} \pi_{\sf Si} ~+~ R^t_{\sf Ai} \pi_{\sf Ai} \right)  \\ &\quad~~~~~~~~~~~~~ \times \Bigg( 1 ~-~  {\tt nonCOVIDSymptRate}   ~-~  {\tt contactPerPosCase} \times  \frac{  \bar{I}^t_{\sf Si} \pi_{\sf Si} + \bar{I}^t_{\sf Ai} \pi_{\sf Ai}  + \bar{I}^t_{\sf An} \pi_{\sf An} }{\bar{H}}   \Bigg) \\
&\quad~~~~ 
~+~ \frac{R_{\sf Ai}^t (1-\pi_{\sf Ai})}{{\tt selfQuarTime}}    
~+~ \frac{R_{\sf Si}^t (1-\pi_{\sf Si})}{{\tt selfQuarTime}}   
~+~ \frac{ \bar{I}^t_{\sf Si} (1 - \pi_{\sf Si})}{{\tt sympToRecoveryTime}}\\
&\quad~~~~ 
~+~ \frac{ \bar{I}^t_{\sf Ai} (1 - \pi_{\sf Ai})}{{\tt asympToRecoveryTime}} 
~+~ \frac{ \bar{I}^t_{\sf An} (1 - \pi_{\sf An})}{{\tt asympToRecoveryTime}}~.
\end{align*}
}}


\vspace{10pt}
\item{{\bf Known Infected Compartment ($\bm{KI}$)}:}
This compartment has two sub-compartments, with    $\bm{KI}^t = \left( {KI}^t(\mathrm{hospitalized}), H^t(\mathrm{recovered}) \right)$ and we have the following dynamics.

\vspace{10pt}
\uline{Inflow}: There are three sources of inflow. The infected individuals in $\bm{I}_{\sf Si}$, $\bm{I}_{\sf Ai}$, and $\bm{I}_{\sf An}$ compartments who are tested. We assume that individuals from the $\bm{I}_{\sf Si}(\mathrm{hospitalized})$ sub-compartment flow into the ${KI}^t(\mathrm{hospitalized})$ sub-compartment, and that individuals from the $\bm{I}_{\sf Si}(\mathrm{recovered})$ sub-compartment flow into the ${KI}^t(\mathrm{recovered})$ sub-compartment. We assume that {\tt hospOutOfSymptFrac} (20\%) of the individuals from the $\bm{I}_{\sf Ai}$ and $\bm{I}_{\sf An}$ compartments require hospitalization, while the remaining 1 - {\tt hospOutOfSymptFrac} = 80\% will recover naturally.

\vspace{10pt}
\uline{Outflow}: Individuals who require hospitalization move to the $\bm{H}$ compartment at the rate of $1/{\tt infToHospTime}$, where {\tt infToHospTime} represents the average time until an infected person requires hospitalization. We currently set this to 5 days.  Individuals who will recover move to the $KR$ compartment at the rate of
$1/{\tt sympToRecoveryTime}$, where {\tt sympToRecoveryTime} denotes the average recovery time. We set this to 14 days.

\vspace{10pt}

\uline{Update Equations}: Let ${\tt recoverOutOfSymptFrac} = 1 - {\tt hospOutOfSymptFrac} = 80\%$. Then,
{\fontsize{10}{10}\selectfont{
\begin{align*}
{KI}^{t+1}(\mathrm{recovered}) &~=~ \bigg({KI}^t(\mathrm{recovered}) ~+~  I^t_{\sf Si}(\textrm{recovered}) \pi_{\sf Si} ~+~ I^t_{\sf Ai}(\textrm{recovered}) \pi_{\sf Ai} ~+~ I^t_{\sf An}(\textrm{recovered})\pi_{\sf An}\\
&~~~~~~~~~~~~~~~~~ +~{\tt recoverOutOfSymptFrac}  \times  \left[ I^t_{\sf Ai}(\mathrm{sympt.}) \pi_{\sf Ai} ~+~  I^t_{\sf An}(\mathrm{sympt.})\pi_{\sf An} \right]  \bigg)\\
&~~~~~~~~~~~~~~~~~~~~~~~~~~\times \left( 1 -  \frac{1}{{\tt sympToRecoveryTime}}  \right)~, \\
{KI}^{t+1}(\mathrm{hospitalized}) &~=~ \bigg(  {KI}^t(\mathrm{hospitalized})  ~~+~   I^t_{\sf Si}(\mathrm{hospitalized}) \pi_{\sf Si}\\
&~~~~~~~~~~~~~~~~~ + {\tt hospOutOfSymptFrac} \times \left[ I^t_{\sf Ai}(\mathrm{sympt.}) \pi_{\sf Ai} \,+\,  I^t_{\sf An}(\mathrm{sympt.})\pi_{\sf An}   \right] \bigg)\\
&~~~~~~~~~~~~~~~~~~~~~~~~~~\times \left( 1 -  \frac{1}{{\tt infToHospTime}}  \right)~.\\
\end{align*}
}}



\vspace{-5pt}
\item{{\bf Hospitalization Compartment ($\bm{H}$)}:}
This compartment has two sub-compartments, with    $\bm{H}^t = \left( {H}^t(\mathrm{die}), H^t(\mathrm{recovered}) \right)$.

\vspace{10pt}
\uline{Inflow}: There are two sources of inflow to the hospitalization compartment: the infected individuals from the $\bm{I}_{\sf Si}$ and $\bm{KI}$ compartments who require hospitalization. We assume that a fraction ${\tt deathFrac}$, currently set at 1/3, of the inflow to the hospitalization compartment will die. We assume the remaining $1- {\tt deathFrac} = 2/3$ will recover.

\vspace{10pt}
\uline{Outflow}: Individuals who die leave the hospital at the rate of $1/{\tt hospToDeathTime}$, where {\tt hospToDeathTime} is the average time between hospitalization and death.
Individuals who will recover move to the $KR$ compartment at the rate of $1/{\tt hospToRecoveryTime}$, where {\tt hospToRecoveryTime} is the average time between hospitalization and recovery. We currently set both to 14 days.

\vspace{10pt}
\uline{Update Equations}: Let ${\tt recFrac} = 1- {\tt deathFrac} = 2/3$.
We have the following update equations.
{\fontsize{10}{10}\selectfont{
\begin{align*}
H^{t+1}(\mathrm{die}) &~=~ H^{t}(\mathrm{die}) \times \left( 1 -  \frac{1}{{\tt hospToDeathTime}}  \right) ~+~ \\
&\quad \quad \frac{ {\tt deathFrac} }{{\tt infToHospTime}}  \times \Bigg(  I^t_{\sf Si}(\mathrm{hosp.})\times (1-\pi_{\sf Si}) ~+~ 
{KI}^t(\mathrm{hosp.}) ~+~ 
I^t_{\sf Si}(\mathrm{hosp.}) \pi_{\sf Si} \\
&\quad \quad ~~~~~~~~~~~~~~~~~~~~~~~  ~+~ {\tt hospOutOfSymptFrac} \times \left[ I^t_{\sf Ai}(\mathrm{sympt.}) \pi_{\sf Ai} ~+~  I^t_{\sf An}(\mathrm{sympt.}) \pi_{\sf An}   \right]  \Bigg)~, \\
H^{t+1}(\mathrm{recovered}) &~=~ H^{t}(\mathrm{recovered}) \times \left( 1 -  \frac{1}{{\tt hospToRecoveryTime}}  \right) ~+~ \\
&\quad \quad \frac{ {\tt recFrac} }{{\tt infToHospTime}}  \times \Bigg(  I^t_{\sf Si}(\mathrm{hosp.})\times (1-\pi_{\sf Si}) ~+~ 
{KI}^t(\mathrm{hosp.}) ~+~ 
I^t_{\sf Si}(\mathrm{hosp.}) \pi_{\sf Si} \\
&\quad \quad ~~~~~~~~~~~~~~~~~~~~~~~ ~+~ {\tt hospOutOfSymptFrac} \times \left[ I^t_{\sf Ai}(\mathrm{sympt.}) \pi_{\sf Ai} ~+~  I^t_{\sf An}(\mathrm{sympt.}) \pi_{\sf An}   \right]  \Bigg)~.
\end{align*}
}}


\vspace{-10pt}
\item{{\bf Known Recovery Compartment ($KR$)}:}

\vspace{10pt}
\uline{Update Equations}: Let ${\tt recoverOutOfSymptFrac} = 1 - {\tt hospOutOfSymptFrac} = 80\%$. We have the following update equations.
\begin{align*}
{KR}^{t+1} ~=~ {KR}^{t} &~+~  \frac{H^{t}(\mathrm{rec.}) }{{\tt hospToRecoveryTime}}  ~+~
\frac{ {KI}^t(\mathrm{rec.}) }{{\tt sympToRecoveryTime}}\\
&~+~
\frac{   I^t_{\sf Si}(\textrm{rec.}) \pi_{\sf Si} ~+~ I^t_{\sf Ai}(\textrm{rec.}) \pi_{\sf Ai} ~+~ I^t_{\sf An}(\textrm{rec.})\pi_{\sf An}}{{\tt sympToRecoveryTime}} \\
&~+~ \frac{ {\tt recoverOutOfSymptFrac} \times \Big[ I^t_{\sf Ai}(\mathrm{sympt.}) \pi_{\sf Ai} ~+~ I^t_{\sf An}(\mathrm{sympt.})\pi_{\sf An} \Big] }{ {\tt sympToRecoveryTime} }~.
\end{align*}



\vspace{10pt}
\item{{\bf Death Compartment ({\sf D})}:}

\vspace{10pt}
\uline{Update Equations}: We have the following update equations.
\begin{align*}
{D}^{t+1} &~=~ {D}^{t} ~+~ \frac{ H^{t}(\mathrm{die}) }{{\tt hospToDeathTime}} + \dfrac{I^t_{\sf Si}(\mathrm{death})}{{\tt sympToDeathTime}}~.
\end{align*}

\end{itemize}

\section{Model Parameters and Their Estimated Values}
\label{section:model-parameters}


\vspace{10pt}
In this section, we discuss how we obtain the estimates for different model parameters.

\noindent \textbf{Estimating Flow Rates Between Compartments:} 

\noindent We assume an incubation period of $5$ days~\cite{Linton2020}. 
After an individual develops symptoms, we assume it takes another $5$ days for the individual to become hospitalized~\cite{Wang2020}. Specifically, we set ${\tt infToSympTime}={\tt infToHospTime} = {\tt sympToHospTime}=5$.

We set the recovery time for symptomatic individuals to be $14$ days. In our model, we do not differentiate between recovery at home versus at the hospital and set ${\tt sympToRecoveryTime} = {\tt hospToRecoveryTime} =14$. We set the time for a symptomatic individual to die to be $14$ days. Similar to the recovery time, we do not differentiate between deaths at home versus at the hospital and set both ${\tt sympToDeathTime}={\tt hospToDeathTime}=14$~\cite{Linton2020}.  

We assume that individuals who self-isolate after being identified as a close-contact or develop symptoms from a non-COVID-19 illness spend an average of $10$ days reducing their contact with others. While the WHO recommends a quarantine period of 14 days, the CDC gives a less stringent recommendation of 10 days~\cite{CDC2020}. We set ${\tt selfQuarTime}={\tt asympToRecoveryTime}=10$. 

In our model, non-infected and recovered individuals develop symptoms due to non-COVID-19 diseases at a rate of ${\tt nonCOVIDSymptRate}=\frac1{1200}$. This rate is based on the assumption that $10\%$ of the population is infected over the course of $4$ months. This is roughly estimated using the flu symptom rate in past years as reported by the CDC~\cite{CDCFlu2020}. We set $N=8,550,971$, the population of NYC in 2020 as projected by the City~\cite{NYCOpenData2020}. We also assume that $28,000$ individuals, corresponding to roughly $3\%$ of the population, initially has symptoms due to the seasonal flu. This is calculated assuming that flu symptoms last for an average of $4$ days.

\noindent \textbf{Estimating Disease Dynamics:} 

\noindent We use ${\tt sympOutOfInfFrac}$ to denote the fraction of COVID-19 infected individuals who develop symptoms. Numerous reports have given different estimates for this fraction: from $82\%$ on the Diamond Princess cruise ship~\cite{MKZC2020} to roughly $40\%$ on the USS Theodore Roosevelt~\cite{Reuters2020}. We set ${\tt sympOutOfInfFrac}=50\%$.

The parameter ${\tt hospOutOfSympFrac}$ denotes the fraction of symptomatic individuals who need hospitalization. We set this value to be $20\%$. The parameter ${\tt deathOutOfHospFrac}$ denotes the fraction of hospitalized individuals who die. We set this value to be $33.\bar{3}\%$. Both of these values are roughly estimated using historical hospitalization and death numbers as reported by New York City \cite{NYCdata}. Finally, we assume that some portion of the infected population die at home without hospitalization. We denote this portion as ${\tt deathOutOfInfFrac}$ and set it to $2\%$. We estimate this value using the number of probable deaths as reported by the City.

\noindent\textbf{Infection Dynamics and Estimating Contact Rates:}

\noindent The basic reproductive number, $R_0$, captures the average number of new infections produced by each infected individual. There are numerous recent studies estimating the value of $R_0$ for COVID-19. These estimates vary widely, from anywhere between 2.2~\cite{Li2020} and 5.7~\cite{Sanche2020}. Furthermore, it is possible that $R_0$ varies across different countries and cities due to climate, environmental, and sociological differences. As a result, we set $R_0=3.38$ according to a tuning procedure using historical hospitalization and death statistics as reported by New York City. We briefly discuss the details of this procedure in Section \ref{section:model-validation}.

The rate of new infections is controlled by $\beta$, the contact rate between infected and non-infected populations. In a standard compartmentalized model, the number of new infections during any time period is captured by 
$$\beta \times \{\# \text{ of non-infected individuals}\} \times \{\# \text{ of infected individuals}\}.$$
In our model, we have a similar term for every pair of infected and non-infected compartments. We employ two different contact rates $\beta_h$ and $\beta_\ell$, so we differentiate their usage according to whether we are considering isolating or non-isolating compartments.

We set $\beta_h=\frac{R_0}{N\times {\tt sympToRecoveryTime}}$ and use this as the contact rate between infected and \mbox{non-infected} compartments that are both non-isolating.
When at least one of the compartments is isolating or symptomatic, we assume that contact between these populations is reduced by one third and use the contact rate $\beta_\ell = \frac23\times \frac{R_0 }{N \times {\tt sympToRecoveryTime}}$.

Beginning on March 22, we reduce both $\beta_h$ and $\beta_\ell$ by a factor of $1/3$. This reduction is due to social distancing measures associated with the start of NY Pause. We set this reduction to $1/3$ according to a report by SafeGraph which used smartphone location data to estimate the reduction in foot traffic due to social distancing measures across the United States. Their study shows that starting March 22, foot traffic in the New York region is at roughly one third of the activity level recorded in previous years over the same time period~\cite{safegraph2020}.

\noindent\textbf{Dynamics of Contact Tracing:}

\noindent The parameter ${\tt contactPerPosCase}$ denotes the average number of close contacts that are identified via tracing for each COVID-19 positive individual. We assume that close contacts are always identified from within the non-isolating population.
The parameter ${\tt likOfBeingInfected}$ is a multiplier that captures how much more likely an infected individual will be identified as a close contact as compared to a non-infected individual.
Specifically,
\[
{\tt likOfBeingInfected} = \frac{P(\mbox{close-contact} \ts | \ts  \mbox{infected})}{P(\mbox{close-contact} \ts | \ts  \mbox{non-infected})},
\]
which can be shown to be equal to 
\[
{\tt likOfBeingInfected} = \frac{P(\mbox{infected} \ts | \ts  \mbox{close-contact})}{P(\mbox{non-infected} \ts | \ts  \mbox{close-contact})} \div \frac{P(\mbox{infected})}{P(\mbox{non-infected})}.
\]

Thus, we need estimates for the proportion of infected versus non-infected individuals within the non-isolating population as well as the proportion of infected versus non-infected individuals among people identified as close contacts. Unfortunately, to the best of our knowledge there are no studies that can be used to estimate these proportions for New York City. Therefore, we use data reported for Wuhan, the capital of Hubei province in China, to estimate these proportions. Wuhan has a population of approximately 11 million people and is comparable in size to New York City. It experienced a surge of COVID-19 cases similar to that of New York City.

To estimate ${\tt contactPerPosCase}$ and the proportion of infected and non-infected individuals within the close-contact group, we use the results of~\cite{Bi2020}, a comprehensive study on contact tracing. This study reports the results of contact tracing in Shenzhen, China but the index cases are mostly travelers arriving from Hubei province. In this study, the authors identified 1,286 close contacts based on 292 COVID-positive cases. This suggests an average close contact group size of $1286 / 292 = 4.4$ for each COVID-positive individual identified. According to recent news articles, the state of Massachusetts, which has recently initiated a comprehensive contact-tracing effort, found that the average close contact group size is two~\cite{nbcb2020}. Therefore, we set the parameter ${\tt contactPerPosCase}$ to 4 in our model. Of the 1,286 close contacts, 98 tested positive. Thus, we estimated the proportion of the infected to non-infected within the close contact group to be $P(\mbox{infected} | \mbox{close-contact})/P(\mbox{non-infected} | \mbox{close-contact}) = 98/1188 = 0.082$. 


To estimate $P(\mbox{infected})/P(\mbox{non-infected})$, the proportion of infected to non-infected in the general non-isolating population, we used the following approach. According to a recent study, in Wuhan, the probability of death after developing symptoms was 1.4\%~\cite{Wu2020}. Using this estimate and the number of deaths in Wuhan, which is reported as 4,512 by the Johns Hopkins Coronavirus Resource Center~\cite{JHCRC2020}, we can estimate the total number of infected individuals in Wuhan to be 322,286. 
This estimates the total number of infected individuals over the entire course of the pandemic in Wuhan. In order to estimate the total number of infected individuals at some random time between mid-January and mid-February, we divide this by two and use 161,143 as the average number of infected individuals at a given time. This approximation implicitly assumes a linear build-up of new COVID-positive cases, which is not true for infection spread within a population, but nevertheless serves as a reasonable rough estimate. Then, using 11,000,000 for the population of Wuhan, we estimate $P(\mbox{infected})/P(\mbox{non-infected})$ to be $161,143/(11,000,000-161,143) = 0.015$.

Finally, using our estimates for the infected to non-infected proportion within the close-contacts and within the general non-isolating population, we estimated the ratio of these two proportions to be $0.082/0.015 = 5.47$ and set ${\tt likOfBeingInfected}$ to 5.5.

\section{Model Validation} \label{section:model-validation}

\vspace{10pt}


We manually calbirated two parameters of the model to roughly match its output to the observed trajectory of the pandemic up until April 15. These two parameters are the $R_0$ value and the initial number of infected asymptomatic-nonisolated individuals on March 2. We ended up with an $R_0$ value of 3.38 and an initial infected asymptomatic-nonisolated population of about 10,000. These starting parameters are tuned based on hospitalization, death, and test data up to April 15. We use testing data from NYS Health department \cite{NYSdata}. We assume that the tests are conducted daily on people in the order of symptomatic, isolated, and finally non-isolated people. 

\begin{figure}[!tbp]
  \centering
  \subfloat
  [Population overview]
  {\includegraphics[scale=0.59]{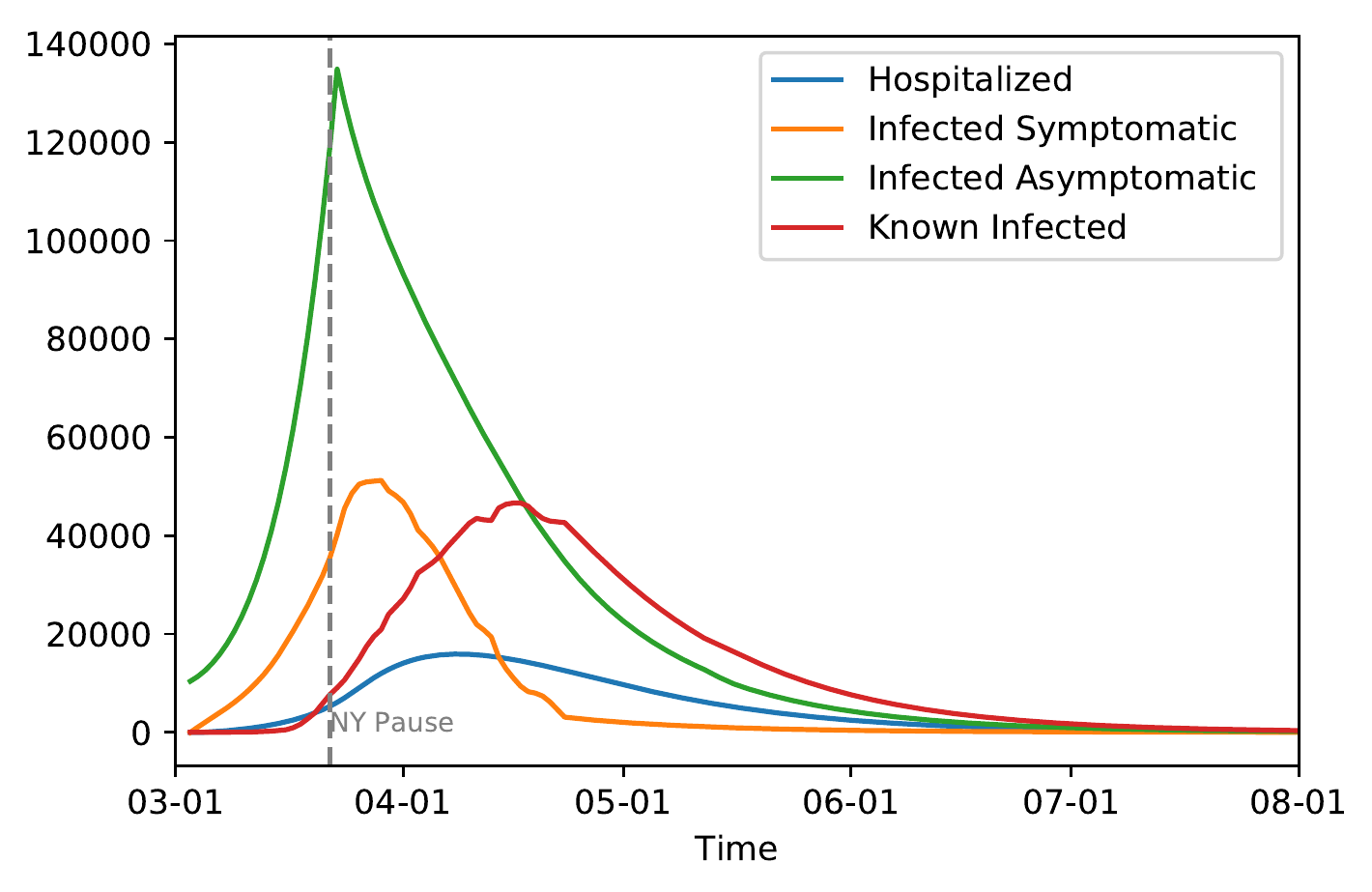}
  \label{fig:overview}}
  \hfill
  \subfloat
  [Cumulative deaths]
  {\includegraphics[scale=0.55]{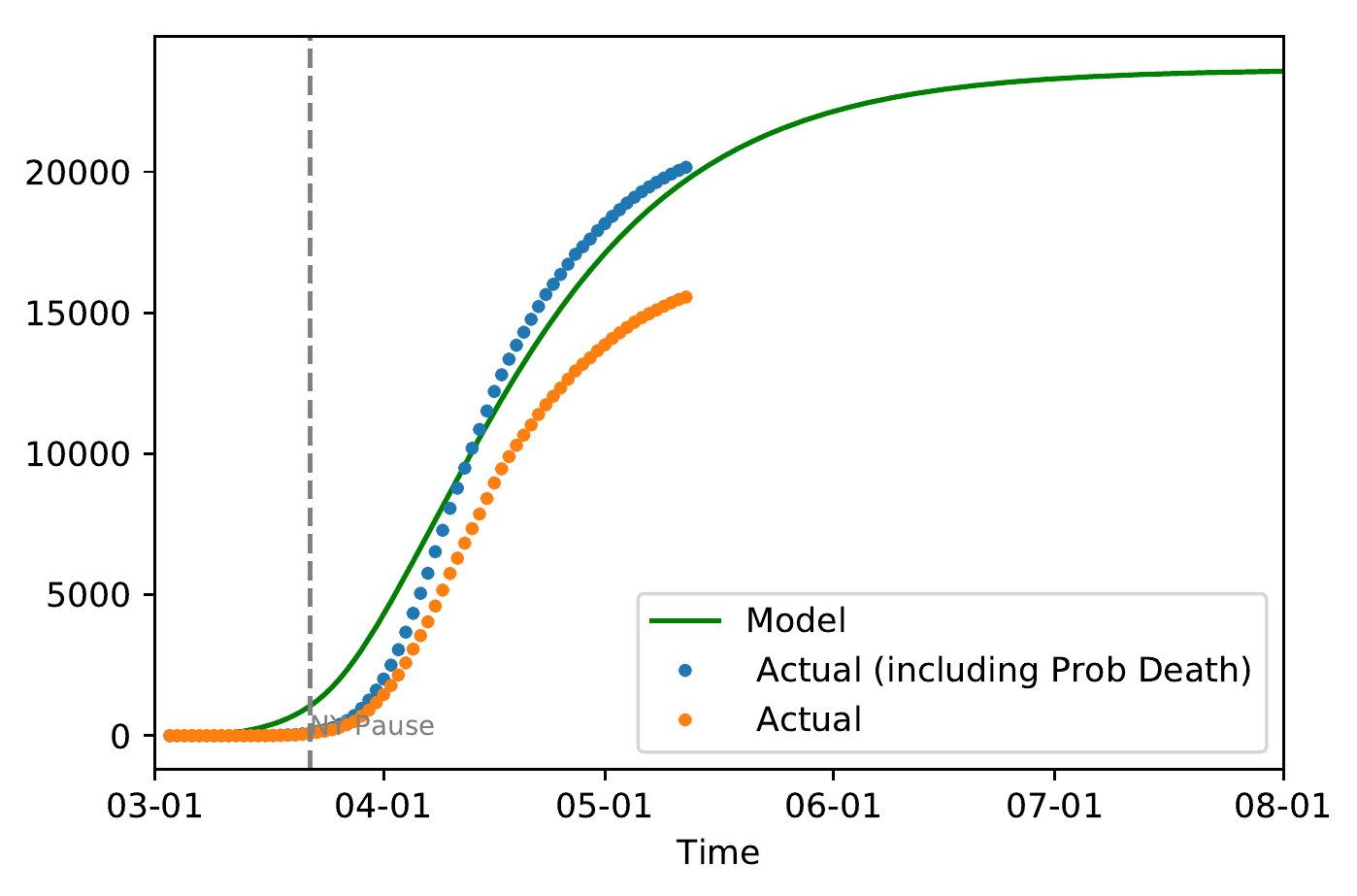}
  \label{fig:cum-deaths}}
  \hfill
  \subfloat
  [Daily new cases]
  {\includegraphics[scale=0.55]{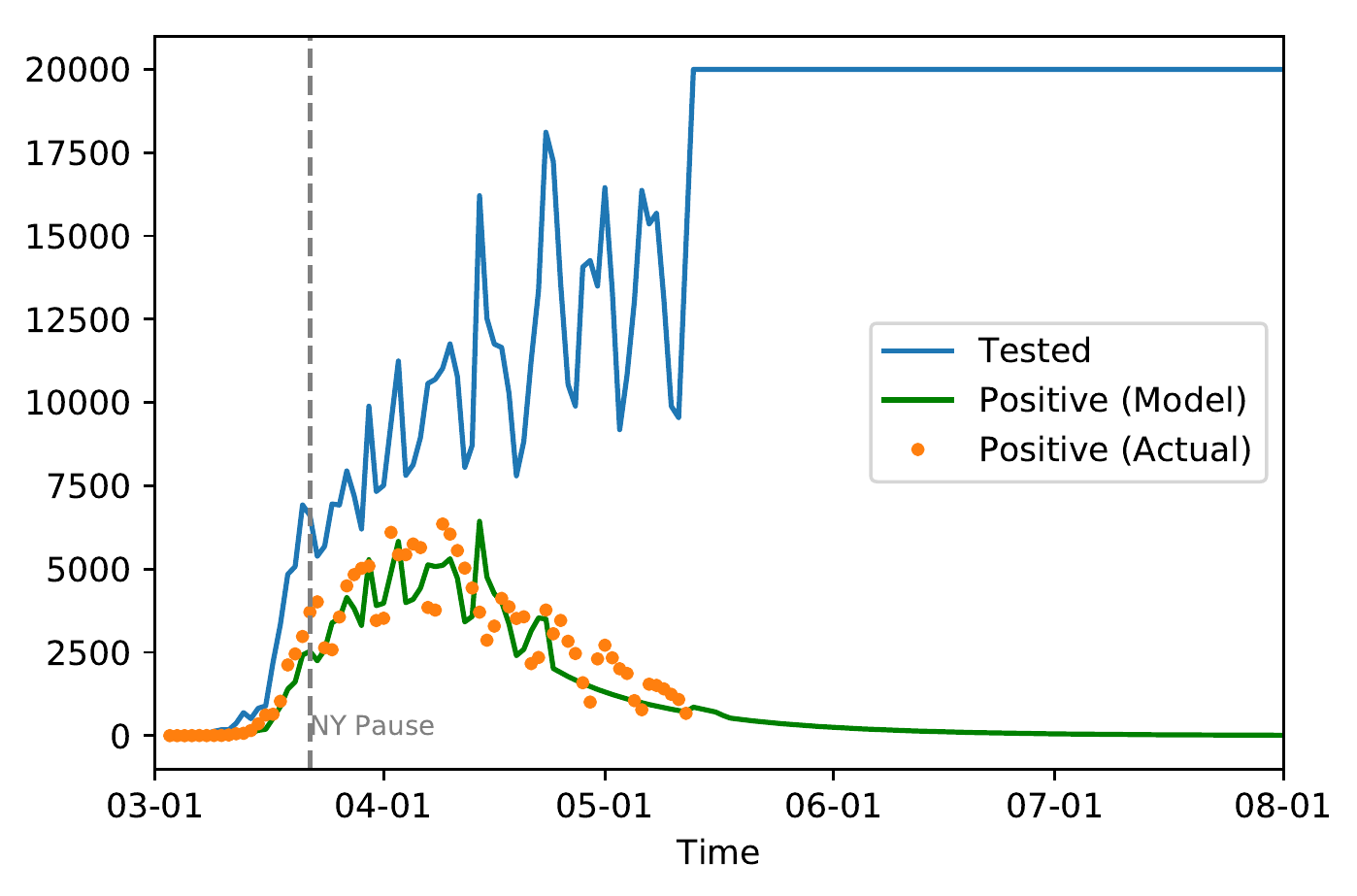}
  \label{fig:cases}}
  \hfill
  \subfloat
  [Daily new hospitalizations]
  {\includegraphics[scale=0.58]{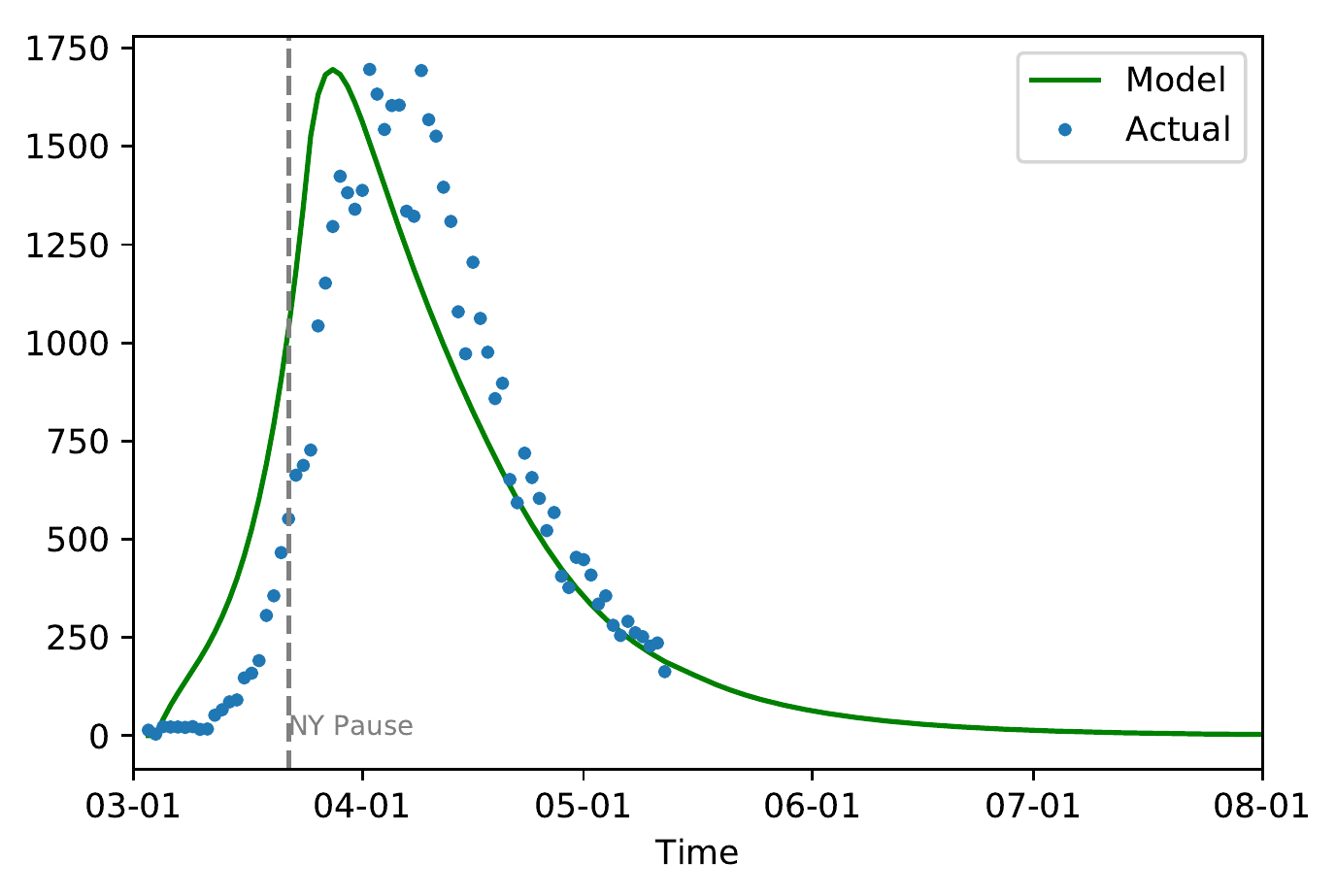}
  \label{fig:new-hosp}}
  \hfill
  \subfloat
  [Daily new deaths]
  {\includegraphics[scale=0.58]{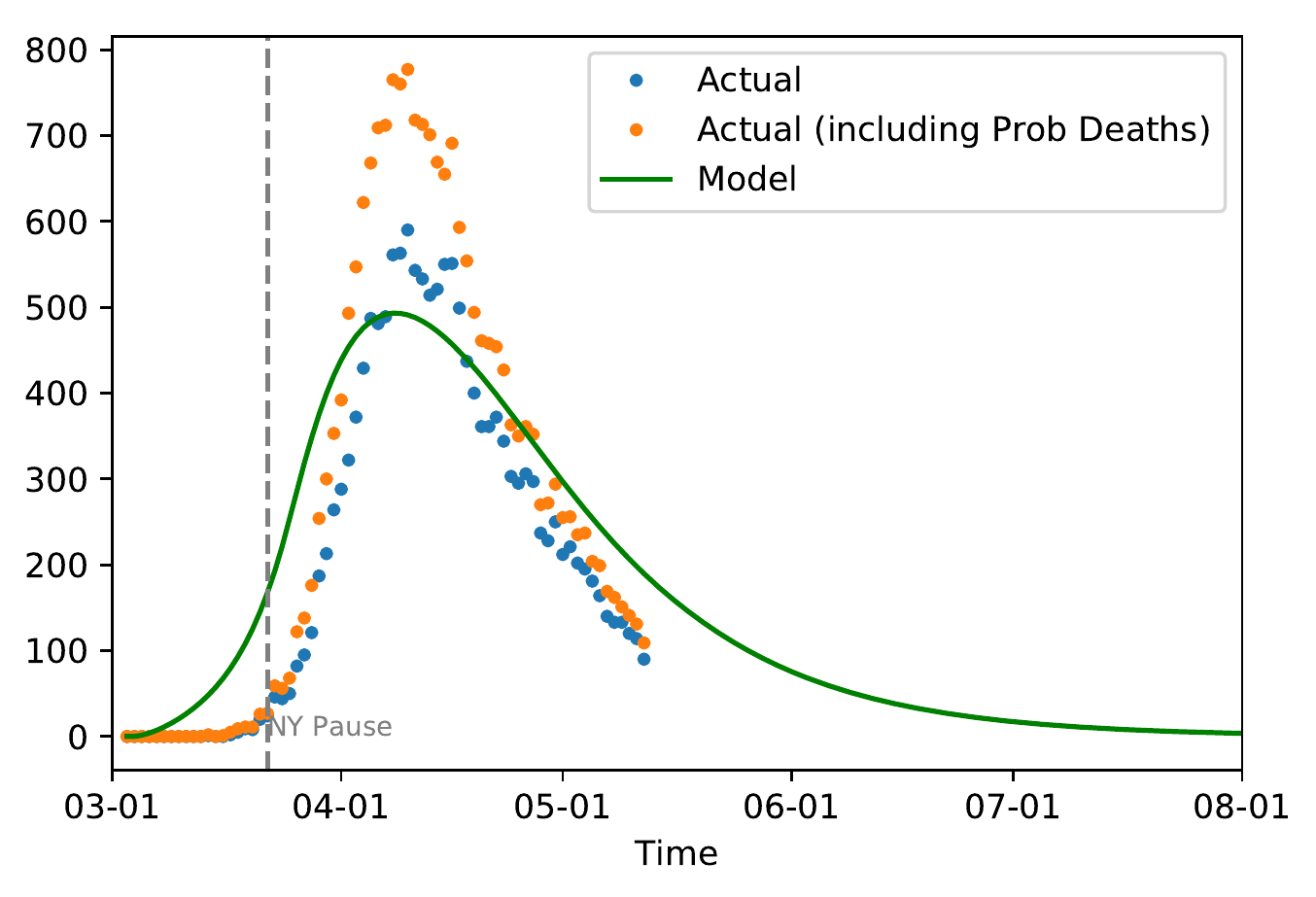}
  \label{fig:new-deaths}}
\caption{Comparison of the actual trajectory of the pandemic and the model.} \label{fig:trajectory}
\end{figure}

The plots in Figure \ref{fig:trajectory} show the epidemic trajectory as predicted by our model. In particular, Figure \ref{fig:overview}  shows the total number of currently hospitalized, symptomatic infected, asymptomatic infected and known infected people on each day as projected by the model. 
In Figure \ref{fig:cum-deaths}, we compare the cumulative number of deaths  predicted by our model with the actual cumulative number of deaths. The data set from NYC Health includes  confirmed and probable deaths due to COVID-19. A death is classified as probable if the decedent was a New York City resident who had no known positive laboratory test for COVID-19 but the death certificate lists as a cause of death ``COVID-19'' or equivalent. We show both confirmed deaths and probable deaths in our plots. In Figure \ref{fig:cases}, we plot the actual daily number of tests conducted and positive cases found in the City. We use the same historical testing capacity in our model. We set the testing capacity to 20,000 in the City for future projections in this plot. Note that the daily number of positive tests in our model fluctuates because we use the actual number of tests performed in
the City until early May as the number of tests performed in our model.  However, the number of positive cases identified in our model emerges from the internal behavior of our model. The fit between the model and the actual trajectory in terms of positive cases is reasonably good.
Figure \ref{fig:new-hosp} shows the daily hospitalizations reported by NYC Health against the number of hospitalizations projected by our model. Figure \ref{fig:new-deaths} plots the daily deaths projected by our model against the daily deaths reported by NYC Health, including probable deaths.

\end{document}